\documentclass[aps,prl,a4paper,10pt,twocolumn,showpacs,floatfix,superscriptaddress,amsmath,preprintnumbers]{revtex4-1}

\usepackage{float}
\usepackage{dcolumn,graphicx,color,booktabs,microtype,afterpage}
\graphicspath{{./}{figure/}{suppl_mater/}}
\usepackage[charter,greekuppercase=italicized]{mathdesign}
\usepackage{sidecap}
\renewcommand{\tablename}{Table}
\makeatletter\renewcommand{\fnum@figure}[1]{\figurename~\thefigure.~}\makeatother
\makeatletter\renewcommand{\fnum@table}[1]{\tablename~\thetable.}\makeatother

\newcount\hh \newcount\mm
\hh=\time \divide\hh by 60
\mm=\hh \multiply\mm by 60 \mm=-\mm
\advance\mm by \time
\def\now{\number\hh:\ifnum\mm<10{}0\fi\number\mm}

\usepackage[colorlinks,plainpages=false,linkcolor=blue,urlcolor=blue,citecolor=blue,pdfpagemode=UseNone,pdfstartview=FitBH]{hyperref}

\newcommand{\tcr}[1]{\textcolor{black}{#1}}
\newcommand{\tcb}[1]{\textcolor{black}{#1}}

\definecolor{darkgreen}{rgb}{0,0.5,0}

\begin{document}

\makeatletter\renewcommand{\ps@plain}{%
\def\@evenhead{\hfill\itshape\rightmark}%
\def\@oddhead{\itshape\leftmark\hfill}%
\renewcommand{\@evenfoot}{\hfill\small{--~\thepage~--}\hfill}%
\renewcommand{\@oddfoot}{\hfill\small{--~\thepage~--}\hfill}%
}\makeatother\pagestyle{plain}


%
%
\title{\tcr{Time-reversal symmetry breaking in Re-based superconductors}}
\author{T.\ Shang}\email[Corresponding authors:\\]{tian.shang@psi.ch}
\affiliation{Laboratory for Multiscale Materials Experiments, Paul Scherrer Institut, Villigen CH-5232, Switzerland}
\affiliation{Swiss Light Source, Paul Scherrer Institut, Villigen CH-5232, Switzerland}
\affiliation{Institute of Condensed Matter Physics, \'Ecole Polytechnique F\'ed\'erale de Lausanne (EPFL), Lausanne CH-1015, Switzerland.}
\author{M.\ Smidman}\email[Corresponding authors:\\]{msmidman@zju.edu.cn}
\affiliation{Center for Correlated Matter and Department of Physics, Zhejiang University, Hangzhou 310058, China}
\author{\tcr{S.\ K.\ Ghosh}}
\affiliation{School of Physical Sciences, University of Kent, Canterbury CT2 7NH, United Kingdom}
\author{C.\ Baines}
\affiliation{Laboratory for Muon-Spin Spectroscopy, Paul Scherrer Institut, CH-5232 Villigen PSI, Switzerland}
\author{L.\ J.\ Chang}
\affiliation{Department of Physics, National Cheng Kung University, Tainan 70101, Taiwan}
\author{D.\ J.\ Gawryluk}
\affiliation{Laboratory for Multiscale Materials Experiments, Paul Scherrer Institut, Villigen CH-5232, Switzerland}
\author{J.~A.~T.~Barker}
\affiliation{Laboratory for Muon-Spin Spectroscopy, Paul Scherrer Institut, CH-5232 Villigen PSI, Switzerland}
%
\author{R.\ P.\ Singh}
\affiliation{Indian Institute of Science Education and Research Bhopal, Bhopal, 462066, India}
\author{D.\ McK.\ Paul}
\affiliation{Physics Department, University of Warwick, Coventry CV4 7AL, United Kingdom}
\author{G.\ Balakrishnan}
\affiliation{Physics Department, University of Warwick, Coventry CV4 7AL, United Kingdom}

\author{E.\ Pomjakushina}
\affiliation{Laboratory for Multiscale Materials Experiments, Paul Scherrer Institut, Villigen CH-5232, Switzerland}
\author{M.\ Shi}
\affiliation{Swiss Light Source, Paul Scherrer Institut, Villigen CH-5232, Switzerland}
\author{M.~Medarde}
\affiliation{Laboratory for Multiscale Materials Experiments, Paul Scherrer Institut, Villigen CH-5232, Switzerland}
\author{A.\ D.\ Hillier}
\affiliation{ISIS Facility, STFC Rutherford Appleton Laboratory, Harwell Science and Innovation Campus, Oxfordshire, OX11 0QX, United Kingdom}
\author{H.\ Q.\ Yuan}
\affiliation{Center for Correlated Matter and Department of Physics, Zhejiang University, Hangzhou 310058, China}
\affiliation{Collaborative Innovation Center of Advanced Microstructures, Nanjing Univeristy, Nanjing 210093, China}
\author{\tcr{J. Quintanilla}}\email[Corresponding authors:\\]{j.quintanilla@kent.ac.uk}
\affiliation{School of Physical Sciences, University of Kent, Canterbury CT2 7NH, United Kingdom}
\author{J.\ Mesot}
\affiliation{Paul Scherrer Institut, CH-5232 Villigen PSI, Switzerland}
\affiliation{Institute of Condensed Matter Physics, \'Ecole Polytechnique F\'ed\'erale de Lausanne (EPFL), Lausanne CH-1015, Switzerland.}
\affiliation{Laboratorium f\"ur Festk\"orperphysik, ETH Z\"urich, CH-8093 Zurich, Switzerland}
\author{T.\ Shiroka}\email[Corresponding authors:\\]{tshiroka@phys.ethz.ch}
\affiliation{Laboratorium f\"ur Festk\"orperphysik, ETH Z\"urich, CH-8093 Zurich, Switzerland}
\affiliation{Paul Scherrer Institut, CH-5232 Villigen PSI, Switzerland}

\begin{abstract}
To trace the origin of time-reversal symmetry breaking (TRSB) in 
Re-based superconductors, 
we performed comparative muon-spin rotation/relaxation ($\mu$SR) studies of superconducting noncentrosymmetric 
Re$_{0.82}$Nb$_{0.18}$ ($T_c = 8.8$\,K) and centrosymmetric Re ($T_c = 2.7$\,K). 
In Re$_{0.82}$Nb$_{0.18}$,  the low\--tem\-per\-a\-ture superfluid density and the electronic 
specific heat evidence a fully-gapped superconducting state, whose enhanced 
gap magnitude and specific-heat discontinuity suggest a moderately strong electron-phonon coupling. In both 
Re$_{0.82}$Nb$_{0.18}$ and pure Re, the spontaneous magnetic fields revealed by zero-field 
$\mu$SR below $T_c$ indicate  time-reversal symmetry breaking and thus unconventional superconductivity. The concomitant occurrence of TRSB in centrosymmetric Re
and noncentrosymmetric Re$T$ ($T$ = transition metal), yet its preservation in the isostructural 
noncentrosymmetric superconductors Mg$_{10}$Ir$_{19}$B$_{16}$ and Nb$_{0.5}$Os$_{0.5}$, 
strongly suggests that \tcr{the local electronic structure of Re is crucial for understanding the TRSB superconducting state in Re and Re$T$. We discuss the superconducting order parameter symmetries that are compatible with the observations.}
\end{abstract}



\maketitle\enlargethispage{3pt}

\vspace{-5pt}
%
Time reversal and spatial inversion are two key symmetries which influence at a fundamental 
level the electron pairing in the superconducting state: on the one hand, a number of unconventional 
superconductors exhibit spontaneous time-reversal symmetry breaking (TRSB) on entering the superconducting state; on the other hand, the absence of inversion 
symmetry above $T_c$ leads to an antisymmetric spin-orbit coupling (SOC), lifting the degeneracy of the conduction-band electrons and potentially giving rise to a mixed-parity 
superconducting state~\cite{Bauer2012,smidman2017}. Some noncentrosymmetric superconductors (NCSC), such as CePt$_3$Si~\cite{bonalde2005CePt3Si}, CeIrSi$_3$\cite{mukuda2008CeIrSi3},                                                                                                                                                                                                                                                                                                                 Li$_2$Pt$_3$B~\cite{yuan2006,nishiyama2007}, and K$_2$Cr$_3$As$_3$~\cite{K2Cr3As3Pen,K2Cr3As3MuSR}, exhibit 
line nodes in the gap, while others such as LaNiC$_2$~\cite{chen2013} and (La,Y)$_2$C$_3$~\cite{kuroiwa2008}, 
show multiple nodeless superconducting gaps. In addition, due to the strong influence of SOC, their upper critical 
field can greatly exceed the Pauli limit, as has been found in CePt$_3$Si~\cite{bauer2004} 
and very recently in (Ta,Nb)Rh$_2$B$_2$~\cite{Carnicom2018}.

\tcr{In general, TRSB below $T_c$ and a lack of spatial-inversion symmetry of the crystal structure are independent events.} 
Yet, in a few cases, such as in LaNiC$_2$~\cite{Hillier2009}, 
La$_7$Ir$_3$~\cite{Barker2015}, and, in particular, in the Re-based compounds 
Re$_6$Zr~\cite{Singh2014}, Re$_6$Hf~\cite{Singh2017}, Re$_{6}$Ti~\cite{Singh2018}, 
and Re$_{24}$Ti$_5$~\cite{Shang2018}, \tcr{TRSB below $T_c$ is concomitant with an existing lack of crystal inversion symmetry.} 
Such an unusually frequent occurrence of TRSB among 
the superconducting Re$T$ binary alloys ($T$ = transition metal) is rather 
puzzling. Its persistence independent of the particular transition metal, points 
to a key role played by Re. To test such a hypothesis, and to \tcr{ascertain the possible relevance of the noncentrosymmetric structure to TRSB in Re-based NCSC}, we proceeded with a twofold study. 
On one hand we synthesized and investigated another Re-based NCSC,  
Re$_{0.82}$\-Nb$_{0.18}$. On the other hand, we considered 
the pure Re metal, also a superconductor, but with a centrosymmetric structure.

A comparative study by means of muon-spin relaxation and rotation ($\mu$SR) allows 
us to address the question of TRSB in  Re-containing compounds. 
The choice of $\mu$SR as the preferred technique for our study 
\tcr{is justified by} its key role 
in detecting TRSB in numerous unconventional superconductors~\cite{Luke1998,aoki2003,Hillier2009,Singh2014, Barker2015,Singh2017,Shang2018,Singh2018} (later confirmed by Kerr effect or bulk magnetisation in the cases of Sr$_2$RuO$_4$, UPt$_3$ and LaNiC$_2$~\cite{Xia2006Oct,Schemm2014,Sumiyama2014}). We report  systematic $\mu$SR studies of Re$_{0.82}$\-Nb$_{0.18}$ 
($T_c = 8.8$\,K) and Re ($T_c = 2.7$\,K), whose bulk superconducting properties
were characterized by magnetic, transport, and thermodynamic measurements. 
The $\mu$SR data show that spontaneous magnetic 
fields appear below the respective transition temperatures, thus 
implying that the superconducting states of both Re$_{0.82}$Nb$_{0.18}$ 
and Re show TRSB and have an unconventional nature. 
\tcr{Since pure Re is centrosymmetric, this implies that the noncentrosymmetric structure is not a requirement for TRSB in these materials.}  

%
Polycrystalline Re$_{0.82}$Nb$_{0.18}$ samples were prepared by arc melting 
Re and Nb metals 
and the same Re powder was used for measurements on elementary Re. The x-ray powder diffraction  measured 
using a Bruker D8 diffractometer, confirmed the $\alpha$-Mn structure of Re$_{0.82}$Nb$_{0.18}$ ($I\overline{4}3m$) and the hcp-Mg structure of Re ($P6_3/mmc$)~\cite{Supple,Werthamer1966, Klimczuk2006, Biswas2011, Tari2003, Barron1999, Waterstrat1977}. Magnetic susceptibility, electrical resistivity, and specific-heat measurements were performed on a 7-T Quantum Design Magnetic Property Measurement System and a 9-T Physical Property Measurement System. The $\mu$SR measurements were carried out on both the MuSR instrument 
of the ISIS pulsed muon source (UK)~\cite{DataDOI}, and the 
GPS and LTF spectrometers of the $\pi$M3 beamline at the 
Paul Scherrer Institut, Villigen, Switzerland.
%
%

The magnetic susceptibility was determined using both field-cooled (FC) and 
zero-field-cooled (ZFC) protocols. 
As shown in Fig.~\ref{fig:superconductivity}(a)-(b), the splitting of 
the two curves is typical of type-II superconductors, 
with the ZFC-susceptibility indicating 
bulk superconductivity with $T_c = 8.8$\,K for Re$_{0.82}$Nb$_{0.18}$ and 2.7\,K for Re. 
The bulk super\-conductivity of Re$_{0.82}$Nb$_{0.18}$ was further confirmed 
by electrical resistivity and specific-heat data~\cite{Supple}. 
To perform transverse-field muon-spin rotation (TF-$\mu$SR) measurements of superconductors, the applied field should exceed the lower critical field $\mu_0H_{c1}$, so that 
the additional field-dis\-tri\-bu\-tion broadening due to the 
flux-line lattice (FLL) 
can be determined from the depolarization of the $\mu$SR asymmetry. To determine $\mu_0H_{c1}$, 
the field-dependent magnetization $M(H)$ was measured at various temperatures 
below $T_c$, as shown in Fig.~\ref{fig:superconductivity}(c) for Re$_{0.82}$Nb$_{0.18}$ [for 
$M(H)$ data of Re see Suppl.\ Mater.]~\cite{Supple}. The derived $\mu_{0}H_{c1}$ values are plotted in 
Fig.~\ref{fig:superconductivity}(d) as a function of temperature. The dashed lines 
are fits to $\mu_{0}H_{c1}(T) = \mu_{0}H_{c1}(0)[1-(T/T_{c})^2]$, which
yield estimates of  lower critical fields of 6.4(1)\,mT and 3.7(2)\,mT in 
Re$_{0.82}$\-Nb$_{0.18}$ and Re, respectively. 

\begin{figure}[tb]
  \centering
  \includegraphics[width=0.48\textwidth]{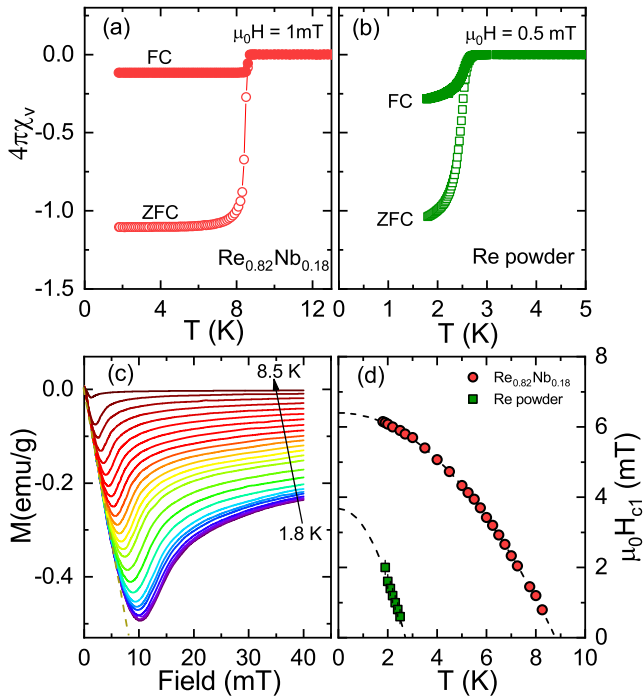}
  \caption{\label{fig:superconductivity} Temperature-dependent  
  magnetic susceptibilities of (a) Re$_{0.82}$Nb$_{0.18}$, and (b) pure Re, measured at 1\,mT and 0.5\,mT, respectively.  
  (c) Magnetization vs.\ applied  magnetic field recorded at different temperatures 
  up to $T_c$ for Re$_{0.82}$Nb$_{0.18}$. For each temperature, $\mu_{0}H_{c1}$ 
  was determined from the value where $M(H)$ deviates from linearity (see dashed line). 
  (d)  $\mu_{0}H_{c1}$ vs.\ temperature for both samples; dashed lines represent
  fits to $\mu_{0}H_{c1}(T) =\mu_{0}H_{c1}(0)[1-(T/T_{c})^2]$.}
\end{figure}
%

The TF-$\mu$SR measurements allowed us to explore the nature of superconductivity 
in Re$_{0.82}$Nb$_{0.18}$ at a microscopic level. The optimal field value for such experiments (above $H_{c1}$) 
was determined via preliminary field-dependent $\mu$SR 
measurements at 1.5\,K~\cite{Supple}. Figure~\ref{fig:TF_MuSR}(a) shows two representative TF-$\mu$SR 
spectra collected  above and below $T_{c}$ in an applied field of 15\,mT. 
In the superconducting mixed state, the faster decay of muon-spin polarization reflects the 
inhomogeneous field distribution due to the FLL. The corresponding TF spectra
are described by: 
\begin{equation}\nonumber
P_\mathrm{TF} = P_\mathrm{s} \cos(\gamma_{\mu} B_\mathrm{s} t + \phi) e^{- \sigma^2 t^2/2} +
P_\mathrm{bg} \cos(\gamma_{\mu} B_\mathrm{bg} t + \phi).
\end{equation}
Here $P_\mathrm{s}$ and $P_\mathrm{bg}$ represent the muon-spin polarization 
for muons implanted in the sample and sample holder,
respectively, with the latter not undergoing any depolarization.
$\gamma_{\mu} = 2\pi \times 135.53$\,MHz/T is the muon gyromagnetic ratio, 
$B_\mathrm{s}$ and $B_\mathrm{bg}$ are the respective local fields
sensed by implanted muons in the sample and sample holder,
$\phi$ is the initial phase, and $\sigma$ is a Gaussian relaxation rate. 
In the superconducting state, the Gaussian relaxation rate includes contributions 
from both the FLL ($\sigma_\mathrm{sc}$) and a temperature-independent relaxation 
due to nuclear moments ($\sigma_\mathrm{n}$). Below $T_c$, 
$\sigma_\mathrm{sc}$ can be extracted after subtracting $\sigma_\mathrm{n}$ in 
quadrature, i.e., $\sigma_\mathrm{sc}$ = $\sqrt{\sigma^{2} - \sigma^{2}_\mathrm{n}}$. 
Since $\sigma_\mathrm{sc}$ is directly related to the magnetic penetration 
depth and hence the superfluid density, the superconducting gap value and its symmetry 
can be determined from the measured relaxation rate. 

As shown in the inset of Fig.~\ref{fig:TF_MuSR}(b), a clear diamagnetic 
shift appears below $T_c$. 
At the same temperature, the formation of the FLL is apparent from the rapid increase of 
$\sigma_\mathrm{sc}$, in turn reflecting an increase of the superfluid density. For small applied magnetic fields [$H_\mathrm{appl}$/$H_{c2}$ $\ll$\,1],
the effective penetration depth $\lambda_\mathrm{eff}$ can be calculated from~\cite{Barford1988,Brandt2003}:
\begin{equation}
\frac{\sigma_\mathrm{sc}^2(T)}{\gamma^2_{\mu}} = 0.00371\, \frac{\Phi_0^2}{\lambda^4_{\mathrm{eff}}(T)}.
\end{equation}
Figure~\ref{fig:TF_MuSR}(b) shows the normalized superfluid density ($[\lambda(T)/\lambda(0)]^{-2}$)
vs.\ the reduced temperature $T/T_c$ for Re$_{0.82}$Nb$_{0.18}$.  
Clearly the temperature dependence of the superfluid density is 
highly consistent between PSI and ISIS, and well described by an $s$-wave 
model with a single gap of 1.61(1)\,meV. By using the 15-mT data,  
\tcr{the resulting $\lambda(0)$ of 357(3)\,nm is comparable with 
352(3)\,nm, the value calculated from $\mu_0H_\mathrm{c1}$~\cite{Supple}}.
The superconducting gap is similar to that of other Re$T$ superconductors, e.g., Re$_6$Zr (1.21\,meV)~\cite{Singh2014, mayoh2017ReZr}, Re$_{24}$Ti$_5$ (1.08\,meV)~\cite{Shang2018}, Re$_6$Ti (0.95\,meV)~\cite{Singh2018}, and Re$_6$Hf (1.10\,meV)~\cite{Singh2017, Singh2016ReHf,chen2016ReHf}, (see also Table SI in Suppl.\ Mater.)~\cite{Supple}. Also the $2\Delta/\mathrm{k}_\mathrm{B}T_{c}$ values of these compounds [e.g., 4.26 for Re$_{0.82}$Nb$_{0.18}$] are higher than 3.53, 
the value expected for weakly-coupled BCS superconductors, thus indicating a moderately strong electron-phonon coupling in these materials. The superconducting parameters of all these $\alpha$-Mn-type Re$T$ NCSC are 
summarized in Table SI in Suppl.\ Mater.~\cite{Supple}.

%
\begin{figure}[thb]
	\centering
	\includegraphics[width=0.4\textwidth,angle=0]{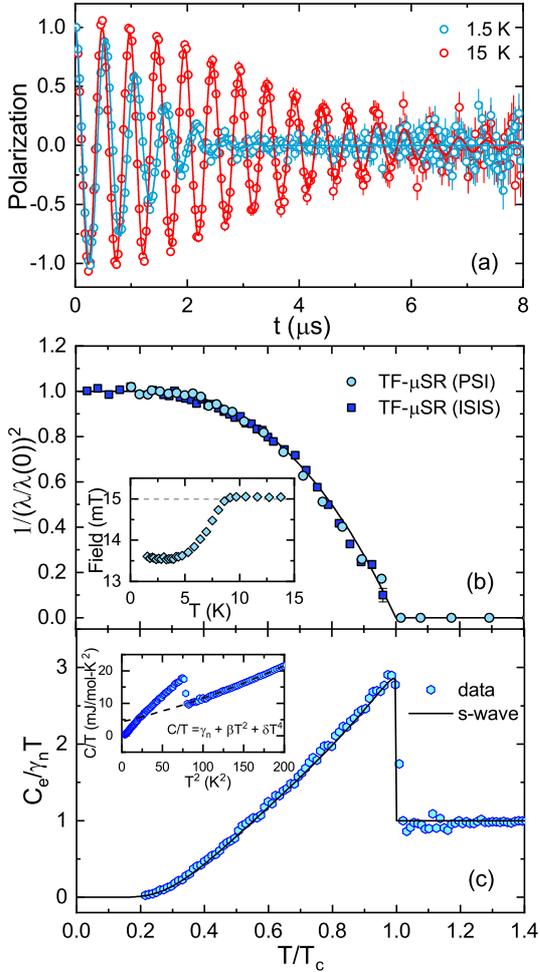}
	\vspace{-2ex}%
	\caption{\label{fig:TF_MuSR} (a) Time-domain TF-$\mu$SR spectra in the superconducting and normal states of Re$_{0.82}$Nb$_{0.18}$, which show very different relaxation rates. (b) Normalized superfluid density vs.\ temperature, as determined from $\mu$SR measurements. Inset: temperature dependence of the internal field across 
	$T_c$. (c) Temperature dependence of the zero-field electronic specific heat. The inset shows the total $C/T$ data versus $T^{2}$. The dashed-line is a fit to $C/T = \gamma_{\mathrm{n}} + \beta T^{2} + \delta T^{4}$, 
	used to estimate the phonon contribution. The solid lines in (b) and (c) represent fits using a fully-gapped $s$-wave model.}
\end{figure}
%

A detailed analysis of the zero-field specific-heat data provides further insight into the superconducting properties of Re$_{0.82}$Nb$_{0.18}$. The electronic specific heat $C_\mathrm{e}$/$T$ was obtained by subtracting the phonon contribution from the experimental data~\cite{Supple}. 
The derived $C_\mathrm{e}/T$ was then divided by the normal-state electronic specific heat coefficient, as shown in the main panel as a function of 
$T/T_c$. The solid line in Fig.~\ref{fig:TF_MuSR}(c) represents a fit with $\gamma_\mathrm{n}$ = 4.4\,mJ\,mol$^{-1}$K$^{-2}$ and a single isotropic gap $\Delta(0)$ = 1.52(2)\,meV. This reproduces very well the experimental data, 
while being consistent with the TF-$\mu$SR [see Fig.~\ref{fig:TF_MuSR}(b)] and previously reported values~\cite{chen2013ReNb, Karki2011}. We also note this value is between the two values found from a two-gap analysis~\cite{Cirillo2015} .
The specific-heat jump at $T_c$ was found to be 
$\Delta C/\gamma_\mathrm{n}T_{c}\sim 1.94$, i.e.,
larger than the conventional BCS value of 1.43, again indicating a moderately enhanced electron-phonon coupling in Re$_{0.82}$Nb$_{0.18}$.

The key goal of the present work is to probe a possible TRSB in 
Re$_{0.82}$Nb$_{0.18}$ and in pure Re. To this aim we performed
detailed zero-field muon-spin relaxation (ZF-$\mu$SR) measurements. 
Normally, in the absence of external fields, the onset of 
superconductivity does not imply changes in the ZF muon-spin relaxation 
rate. However, in presence of TRSB, the onset of a tiny spontaneous polarization 
or currents gives rise to associated (weak) magnetic fields, readily detected by ZF-$\mu$SR as an increase in 
muon-spin relaxation rate. Given the tiny size of such effects, we measured the ZF-$\mu$SR both above $T_c$ and inside the superconducting phase. Representative ZF-$\mu$SR spectra collected above and below $T_c$ for Re$_{0.82}$Nb$_{0.18}$ and Re show measurable
differences [see Fig.~\ref{fig:ZF_muSR}(a)-(b)]. 
To exclude the presence of stray magnetic fields, all 
magnets were quenched before the measurements and an active compensation 
system was used. In non-magnetic materials in the absence of applied fields, 
the relaxation is determined primarily by the randomly oriented nuclear 
dipole moments, normally described by a Gaussian Kubo-Toyabe relaxation 
function~\cite{Kubo1967,Yaouanc2011}. In our case, the ZF-$\mu$SR spectra 
are well described by a combined Lorentzian and Gaussian Kubo-Toyabe relaxation function:
\begin{equation}
\label{eq:KT_and_electr}
P_\mathrm{CKT} = P_\mathrm{s}\left[\frac{1}{3} + \frac{2}{3}(1 -
\sigma^{2}t^{2} - \Lambda t)\,
\mathrm{e}^{\left(-\frac{\sigma^{2}t^{2}}{2} - \Lambda t\right)} \right] + P_\mathrm{bg}.
\end{equation}
Here $P_\mathrm{s}$ and $P_\mathrm{bg}$ are the same as in the 
TF-$\mu$SR spectra.
As shown in the insets of Fig.~\ref{fig:ZF_muSR}, despite the different 
$T_c$ values of Re$_{0.82}$Nb$_{0.18}$ and Re, their $\sigma(T)$ curves 
exhibit a small yet \emph{distinct increase} below $T_c$, similar to that found 
also in other Re$T$ NCSC~\cite{Singh2014,Singh2017,Singh2018,Shang2018}.
At the same time, the Lorentzian relaxation rate $\Lambda(T)$ remains 
mostly constant in the studied temperature range, with typical values 
of 0.007 and 0.005\,$\mu$s$^{-1}$ for Re$_{0.82}$Nb$_{0.18}$ and Re~\cite{Supple}, 
respectively, indicating that fast-fluctuation effects are absent in these systems.
The small, yet measurable increases of $\sigma(T)$ below $T_{c}$, 
detected from measurements at both facilities, reflect the onset of 
spontaneous magnetic fields and thus are signatures of TRSB 
in the superconducting phases of both Re$_{0.82}$Nb$_{0.18}$ and Re. 
\tcr{Further refinements performed by fixing the $\Lambda$ values gave similarly robust features in $\sigma(T)$~\cite{Supple}}. 
To rule out the possibility of a defect-/impurity-induced relaxation 
at low temperatures, we performed auxiliary longitudinal-field $\mu$SR measurements 
at 1.5\,K. As shown in Figs.~\ref{fig:ZF_muSR}(a)-(b), a small field of 15\,mT is sufficient to decouple the muon spins from 
the weak spontaneous magnetic fields in both samples, indicating that the relevant 
fields are static on the time scale of the muon lifetime.

%
\begin{figure}[ht]
	\centering
	\includegraphics[width=0.42\textwidth,angle=0]{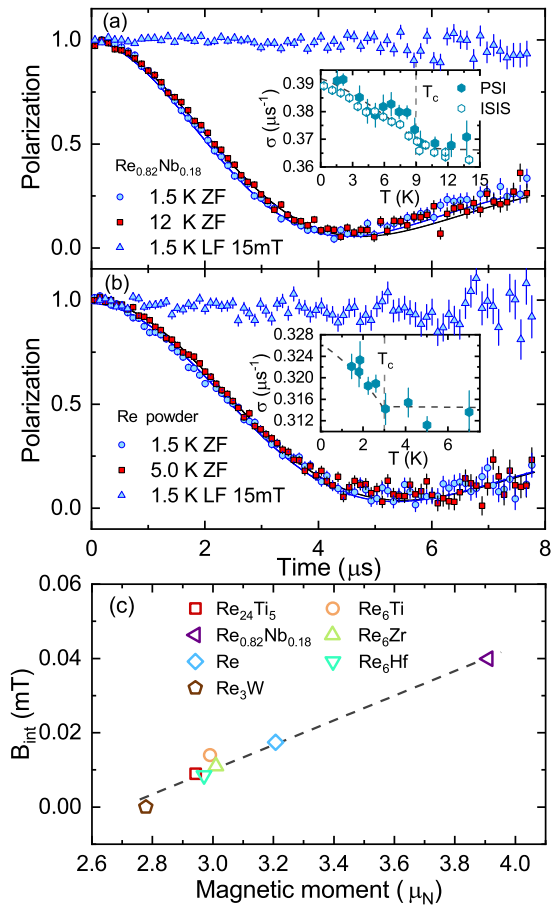}
	\vspace{-2ex}%
	\caption{\label{fig:ZF_muSR} Representative zero-field $\mu$SR spectra for (a) Re$_{0.82}$Nb$_{0.18}$
	and (b) pure Re metal in the superconducting and normal states. Additional $\mu$SR data sets collected 
	at 1.5\,K in a 15-mT longitudinal field, are also shown. The solid lines are fits using Eq.~(\ref{eq:KT_and_electr}). 
	Insets show the $T$-dependence of the relaxation rate $\sigma$. The results from ISIS in (a) were 
	obtained by fitting the ZF data with the corresponding expression in Ref.~\onlinecite{Singh2014}. 
	(c) Calculated internal field 
	vs.\ nuclear magnetic moment for Re$T$ superconductors. For the pure Re (diamond), the value 
	was obtained by an extrapolation to 0\,K. The reported data are 
	from Refs.~\onlinecite{Shang2018,Singh2017,Singh2018,Singh2014,mayoh2017ReZr,Singh2016ReHf,chen2016ReHf,Biswas2012}. The dashed-line indicates a linear behavior. \tcr{The nuclear moment was estimated from the respective nuclear moments $\mu_{n,{\rm Re}}$, $\mu_{n,{\rm T}}$ and chemical fractions $f_{\rm Re}$ and $f_{\rm T}$ of Re and $T$ using $\mu_n = 
        \sqrt{
                f_{\rm Re} \mu_{n,{\rm Re}}^2 + f_{\rm T} \mu_{n,{\rm T}}^2
            }$.}}
\end{figure}
%

To date, most $\alpha$-Mn-type Re$T$-NCSC have been found to exhibit TRSB in the 
superconducting state~\cite{Singh2014,Singh2017,Singh2018,Shang2018}.
Our results show that Re$_{0.82}$Nb$_{0.18}$ is not just 
another member of the Re$T$-NCSC family,
but one with the most distinct TRSB 
in the superconducting state (i.e., with the highest $\sigma_\mathrm{int}$, 
which represents the change of muon relaxation rate between the normal 
and superconducting states (see details in Table SI in Suppl.\ Mater.)~\cite{Supple}. 
This is clearly depicted in Fig.~\ref{fig:ZF_muSR}(c), where we plot 
the estimated internal field $B_\mathrm{int}$ as a function of the 
nuclear magnetic moment $\mu_\mathrm{n}$ 
 for the Re$T$-NCSC.  It can be seen that, $B_\mathrm{int}$ ($\propto \sigma_\mathrm{int}$) scales linearly with $\mu_\mathrm{n}$, 
reaching 0.038\,mT for Re$_{0.82}$Nb$_{0.18}$. 
At the other extreme, \tcr{the line crosses the horizontal axis at $\mu_\mathrm{n} \sim 2.7 \mu_\mathrm{N}$,}
where $\sigma_\mathrm{int}$ drops below the resolution of the $\mu$SR technique (0.01\,mT). 
This is exactly the case for Re$_3$W, whose $\sigma_\mathrm{int}$ 
turned out to be neglible~\cite{Biswas2012}.

\tcr{Having detected TRSB in Re-based superconductors still leaves open  
the most intriguing question: what is its key ingredient? 
In Re$T$, the replacement of heavy 5$d$ elements, such as Hf, 
with lighter 3$d$ elements, such as Ti, appears to have a negligible 
effect on TRSB. The insensitivity of TRSB to the specific transition-metal element $T$ suggests 
that a substitution at the $T$-sites does not significantly influence it. 
This is confirmed by the persistence of TRSB in elemental Re. In addition, this indicates also that a lack of inversion symmetry is inessential. 
Finally, there is no TRSB in the superconducting states of 
Mg$_{10}$Ir$_{19}$B$_{16}$~\cite{Acze2010} and 
Nb$_{0.5}$Os$_{0.5}$~\cite{SinghNbOs}, two NCSCs 
isostructural to Re$T$ and with similar SOC strengths. The above considerations strongly suggest that it is the 
\emph{local electronic structure} of Re that is crucial for 
understanding the TRSB in the superconducting states of Re and Re$T$. 
To reinforce the above conclusion, one could study other Re-free 
materials with the $\alpha$-Mn-type structure. TaOs, 
with a bulk $T_c$ of 2.07\,K and the required crystal structure, 
represents a good example~\cite{SinghTaOs}.
}

\tcr{We now discuss the possible symmetries of the superconducting order parameter. 
In the limit of weak SOC, TRSB can be achieved {\it via} nonunitary 
triplet pairing, as e.g., in LaNiC$_2$~\cite{Quintanilla2010} and 
LaNiGa$_2$~\cite{Hillier2012,Weng2016}. More generally, the relationship between 
TRSB and triplet pairing is quite complex. For example, the admixed 
triplet component in so-called `s-wave' NCSC (those whose superconducting 
instabilities do not break crystal point-group symmetries), 
such as Li$_2$Pt$_3$B, does not show TRSB~\cite{yuan2006}. Conversely, 
there are TRSB states not involving triplet pairing, e.g., the $s+id$ singlet state proposed for some iron-based 
superconductors~\cite{Lee2009}. Apart from the weak-SOC nonunitary triplet pairing scenario mentioned 
above~\cite{Quintanilla2010,Hillier2012}, the essential requirement 
for TRSB occurrence is that the point group of the crystal has 
irreducible representations (irreps) with dimension $D>1$.
}

\tcr{
The point groups $T_d$ and $D_{6h}$ relevant to Re$T$ and Re, 
respectively, have several irreps with $D = 2$ or 3. Therefore, 
they  can support TRSB states  
with singlet-, triplet-, or, in the case of  Re$T$, admixed pairing,
independent of 
SOC strength. In what follows, we will assume strong SOC. 
The full symmetry analysis and plots of the possible order parameters 
can be found in the Suppl. Mat.~\cite{Supple}. For Re$T$ there are 
a number of possible TRSB states, with some examples of pairing 
functions being given in Ref.~\onlinecite{Singh2014}. 
However, all the possible states have symmetry-constrained point or line nodes, inconsistent with the experimental observations. In view of 
this, in some systems, it has been proposed that a full gap 
may be obtained through a Loop-Josephson-Current (LJC) state built 
on on-site, intra-orbital, singlet pairing. Although it has been 
shown that the crystal symmetry of Re$T$ is compatible with this 
scenario~\cite{Ghosh2018}, the energetics which would drive such a state, if realized, 
and why it would occur only in systems with Re and not other 
elements, remain unclear.  
}

\tcr{
The symmetry analysis of 
pure Re contrasts strongly that of Re$T$. Firstly, Re is 
centrosymmetric, implying that the superconducting instability 
can only take place in either purely-singlet or purely-triplet 
channels (irrespective of the strength of SOC). Secondly, due to only two distinct symmetry-related sites 
per unit cell, a LJC state here is not the most natural 
one~\cite{Ghosh2018}. Thirdly, the crystallographic space group is nonsymmorphic, which in principle allows superconducting instabilities that break screw-axis or glide-plane symmetries. Ignoring the ones that break those symmetries, we find two possible TRSB states, one in the singlet channel with a 
line node at $k_z=0$, and another in the triplet channel, with two 
point nodes on the $k_z$ axis (see Suppl.\ Mater.~\cite{Supple}). 
The Fermi surface of Re has five sheets, including an electron 
sheet centered on the $\Gamma$ point and open along the $k_z$ axis and three hole sheets which 
are closed, centered on the $L$ point, and not intersecting the $k_z=0$ plane~\cite{Mattheiss1966}. 
These would be compatible with a full gap for the triplet and singlet TRSB states, respectively.
}
 
In summary, we perfomed comparative $\mu$SR studies of the noncentrosymmetric 
Re$_{0.82}$Nb$_{0.18}$ and centrosymmetric Re superconductors. Bulk 
superconductivity with $T_c = 8.8$\,K (Re$_{0.82}$Nb$_{0.18}$) and 2.7\,K 
(Re) was characterized by magnetic and transport properties. Both the 
superfluid density and the zero-field specific-heat data reveal a 
single-gap nodeless superconductivity in Re$_{0.82}$Nb$_{0.18}$.
The spontaneous fields appearing below $T_c$,  
which increase with decreasing temperature, provide strong evidence that the 
superconducting states of both noncentrosymmetric Re$_{0.82}$Nb$_{0.18}$ 
and centrosymmetric Re show TRSB and are unconventional. 
Comparisons with other Re-free $\alpha$-Mn-type superconductors 
suggest that in the Re$T$ family, the TRSB is crucially
related to the presence of Re, a key idea 
for understanding the peculiar behavior of Re$T$ superconductors.  We have considered 
the possible symmetries of the order parameter in these systems 
and their compatibility with the observed fully-gapped spectrum. 
Further theoretical and experimental work on Re is required 
to clarify the open issues.
  
\textbf{Acknowledgments:} This work was supported by the National Key R\&D Program of China (Grants No.\ 
2017\-YFA\-0303\-100 and 2016\-YFA\-0300\-202), the 
Natural Science Foundation of China (Grant No.\ 11474251), and 
the Schwei\-ze\-rische Na\-ti\-o\-nal\-fonds zur F\"{o}r\-de\-rung
der Wis\-sen\-schaft\-lich\-en For\-schung (SNF). 
The work at the University of Warwick was supported by EPSRC UK 
through Grant EP/M028771/1. Experiments at the ISIS Pulsed Neutron 
and Muon Source were supported by a beamtime allocation from STFC. SKG and JQ are supported by EPSRC through the project ``Unconventional Superconductors: New paradigms for new materials'' (grant references EP/P00749X/1 and EP/P007392/1).
\vspace{-8mm}
\bibliography{ReNb_bib}

%
%

\pagebreak
\clearpage 
\setcounter{equation}{0}
\setcounter{figure}{0}
\setcounter{table}{0}
\setcounter{page}{1}
\makeatletter

\begin{widetext}
\begin{center}
\textbf{\large Supplementary material to\\Time-reversal symmetry breaking in Re superconductors}\\[2mm]
T. Shang,$^{1,2,3}$ M. Smidman,$^4$, S. K. Ghosh,$^5$ C. Baines,$^6$ L. J. Chang,$^7$ D. J. Gawryluk,$^1$ J. A. T. Barker,$^6$ R. P. Singh,$^8$ 
D.~McK.~Paul,$^9$ G.~Balakrishnan,$^9$ E. Pomjakushina,$^1$ M. Shi,$^2$ M. Medarde,$^1$ A. D. Hillier,$^{10}$ H.~Q.~Yuan,$^{4,11}$ 
J.~Quintanilla,$^5$ J.~Mesot,$^{12,3,13}$ and T.~Shiroka$^{13,12}$
\end{center}
\end{widetext}




%
\renewcommand\floatpagefraction{.999}
\renewcommand\topfraction{.999}
\renewcommand\bottomfraction{.999}
	
\maketitle\enlargethispage{3pt}


\renewcommand{\figurename}{FIG.\ S\!\!}
\renewcommand{\tablename}{Table \ S\!\!}

\setcounter{figure}{0}  
\subsection{Crystal structure}  
\begin{figure}[!bht]
\centering
\includegraphics[width=0.75\columnwidth]{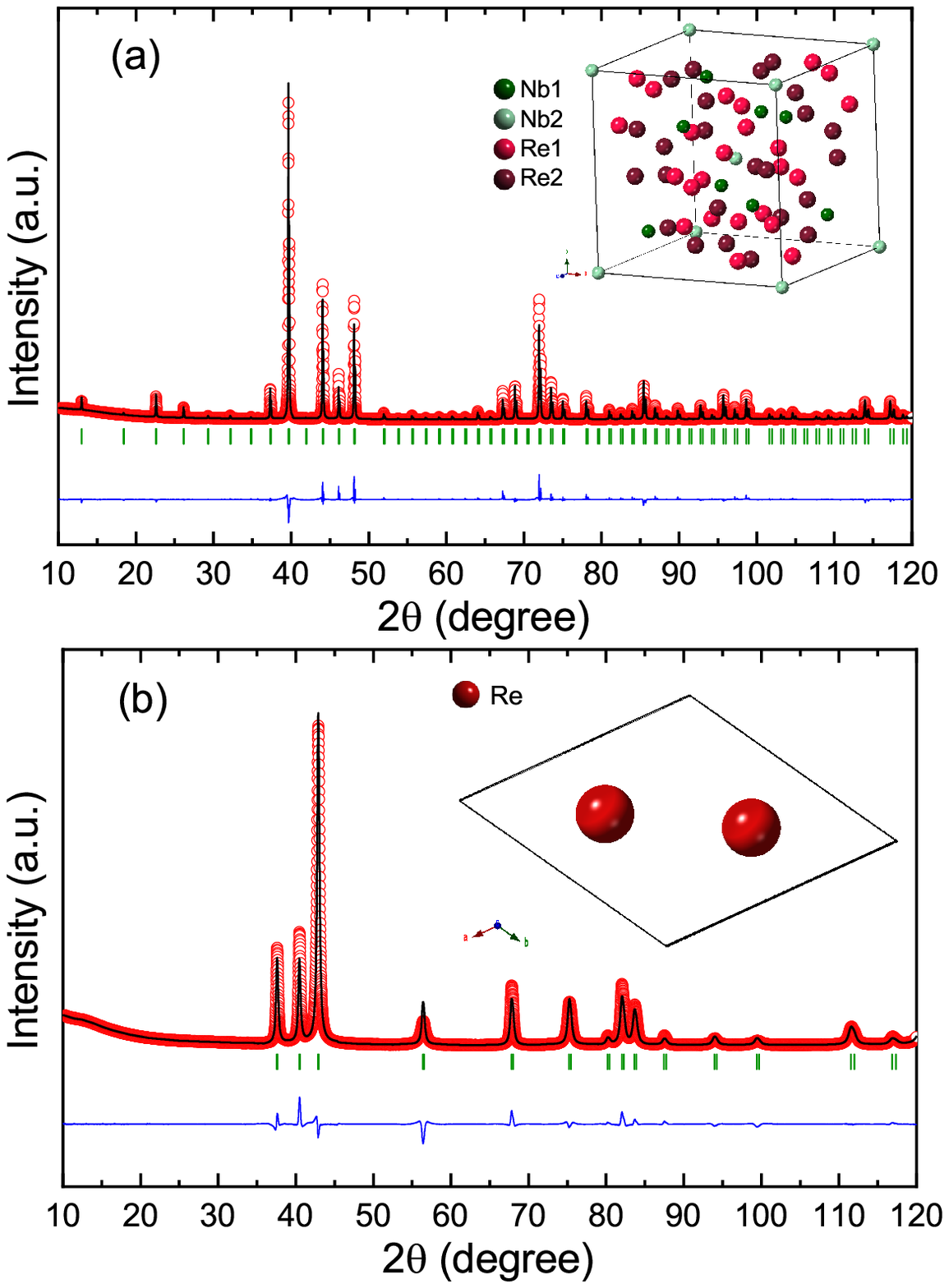} 
\caption{\label{fig:diffraction}Rietveld fits of the room-temperature x-ray powder dif\-frac\-tion 
patterns of Re$_{0.82}$Nb$_{0.18}$ (a) and pure Re (b). Red circles show the experimental 
data, while black lines the refined profiles. Ticks below the patterns indicate the 
Bragg-peak reflections, while blue lines show the residues. Insets show the crystal structures.}
\end{figure}

The purity and crystal structure of both Re$_{0.82}$Nb$_{0.18}$ and pure Re 
were checked via x-ray powder diffraction (XRD). No impurity phases could 
be detected in either sample.  As shown in Fig.~S\ref{fig:diffraction}(a), similarly 
to other Re$T$ alloys, the XRD patterns of Re$_{0.82}$Nb$_{0.18}$ can be well 
indexed by an $\alpha$-Mn-type noncentrosymmetric structure with space 
group $I\overline{4}3m$ (217). On the other hand, as plotted in Fig.~S\ref{fig:diffraction}(b), 
the XRD of pure Re indicates a centrosymmetric hexagonal structure with space 
group $P6_3/mmc$ (194). Both crystal structures are plotted in the insets. 
The derived lattice parameters for pure Re are $a = b = 2.762$\,\AA, $c = 4.457$\,\AA, 
and for Re$_{0.82}$Nb$_{0.18}$ are $a = b = c = 9.636$\,\AA, respectively. 
Table~S\ref{tab:table1} summarizes the atomic coordinates of Re, 
Re$_{0.82}$Nb$_{0.18}$, Nb$_{0.5}$Os$_{0.5}$, and Mg$_{10}$Ir$_{19}$B$_{16}$. 
Also Nb$_{0.5}$Os$_{0.5}$ and Mg$_{10}$Ir$_{19}$B$_{16}$ exhibit an 
$\alpha$-Mn-type noncentrosymmetric structure. Since the latter 
have the same noncentrosymmetric cubic space group as Re$T$, 
it is reasonable to expect a similar SOC-induced band splitting.


\begin{figure}[!tb]
\centering
\includegraphics[width=0.75\columnwidth]{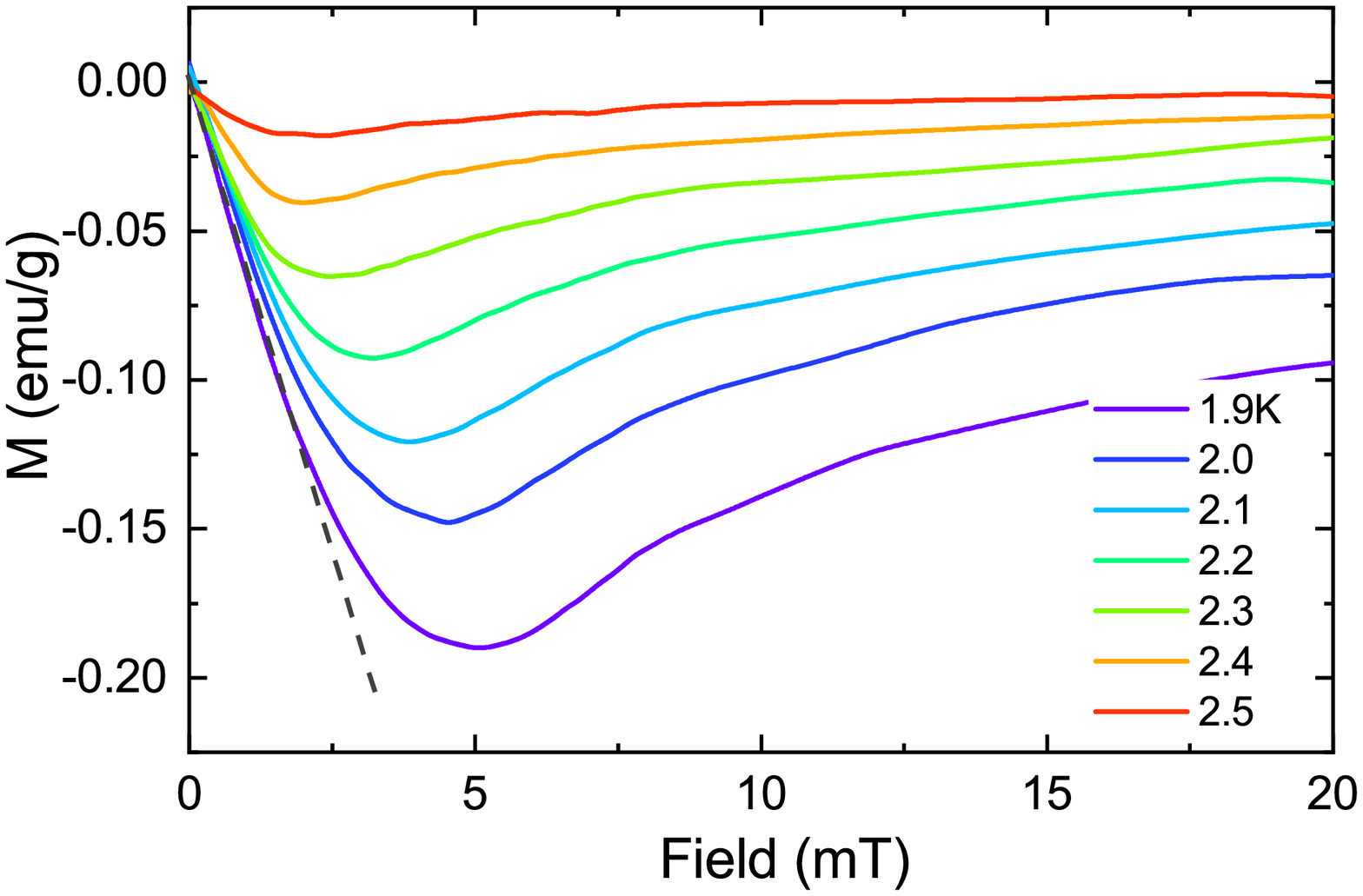} 
\caption{\label{fig:magnetization}Re magnetization vs.\ magnetic field recorded at different 
temperatures up to $T_{c}$. For each temperature, $\mu_{0}H_{c1}$ was determined from 
the $M(H)$ deviation from linearity (dashed line).}
\end{figure}
\subsection{Upper critical field}

\begin{figure}[!htb]
\centering
\includegraphics[width=0.85\columnwidth]{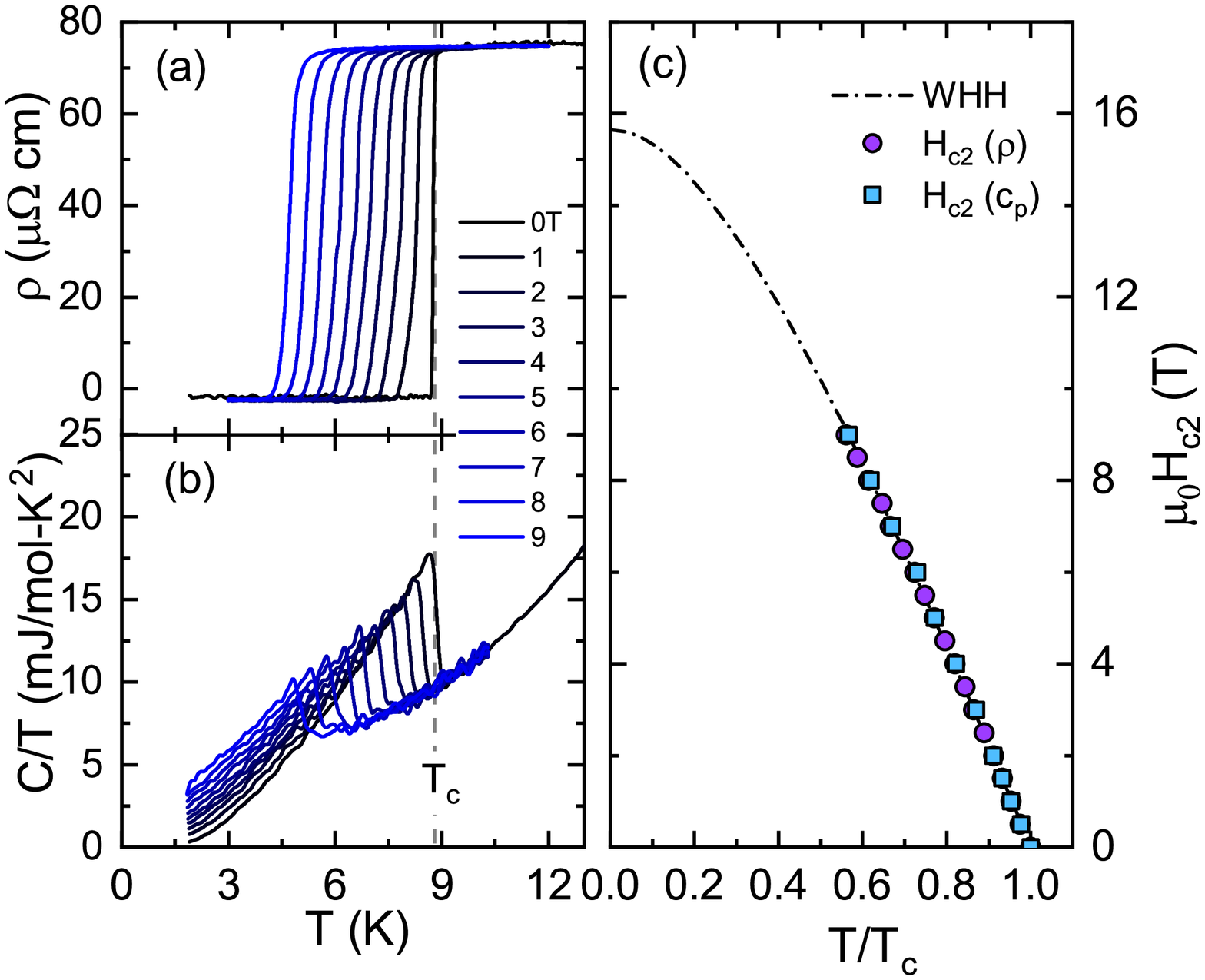} 
\caption{\label{fig:Hc2}Temperature dependence of the (a) electrical 
resistivity, and (b) specific heat, of Re$_{0.82}$Nb$_{0.18}$ in various 
applied magnetic fields up to 9\,T. (c) Upper critical field 
$\mu_{0}H_{c2}$ versus normalized temperature  $T/T_{c}$. The dash-dotted line represents a fit to the WHH model taking into account only the orbital-limiting effect.
}
\end{figure}

As in the Re$_{0.82}$Nb$_{0.18}$ case, also for pure Re we measured the 
field-dependent magnetization $M(H)$   at various temperatures. 
As shown in Fig.~S\ref{fig:magnetization}, $M(H)$ data between 1.9 and 2.5\,K, 
display typical features of type-II superconductors.  
To calculate the penetration depth from $\mu_{0}H_{c1}$, we investigated 
also the upper critical field $\mu_{0}H_{c2}$ of Re$_{0.82}$Nb$_{0.18}$ 
by measuring  its electrical resistivity $\rho(T)$ and 
specific heat 
$C(T)$/$T$ at various magnetic fields up to 9\,T. As shown in 
Fig.~S\ref{fig:Hc2}(a) and (b), both $\rho(T)$ and $C(T)$/$T$ detect 
a zero-field superconducting transition temperature $T_c(0) = 8.8$\,K, 
decreasing with increasing field. The derived upper-critical-field 
$\mu_0H_{c2}(T)$ vs.\ normalized-temperature $T/T_{c}$ data, 
summarized in Fig.~S\ref{fig:Hc2}(c), were further analyzed using
the Werthamer-Helfand-Hohenberg (WHH) model~\cite{Werthamer1966}. The 
dash-dotted line in Fig.~S\ref{fig:Hc2}(c), refers to a WHH model with %
an orbital-limiting field only, which gives 
$\mu_0 H_{c2}^\mathrm{WHH}(0) = 15.6(1)$\,T. This is consistent with 
previously reported values \cite{Karki2011} and very close to the 
weak-coupling Pauli paramagnetic limit  [$\mu_0 H_\mathrm{P} = 1.86\,T_{c} = 16.3(1)$\,T].

In the Ginzburg-Landau theory of superconductivity, the magnetic penetration 
depth $\lambda$ is related to the coherence length $\xi$, and the lower 
critical field via $\mu_{0}H_{c1} = (\Phi_0 /4 \pi \lambda^2)[$ln$(\kappa)+\alpha(\kappa)]$, 
where $\Phi_0 = 2.07 \times 10^{-3}$\,T~$\mu$m$^{2}$ is the quantum of 
magnetic flux, $\kappa$ = $\lambda$/$\xi$ is the Ginzburg-Landau parameter, 
and $\alpha(\kappa)$ is a parameter which converges to 0.497 for $\kappa$ $\gg$ 1~\cite{Brandt2003}. 
By using $\mu_{0}H_{c1} = 6.4$\,mT and $\xi = 4.59$\,nm (calculated from 
$\mu_{0}H_{c2}$) for Re$_{0.82}$Nb$_{0.18}$, the resulting $\lambda_\mathrm{GL} = 352(3)$\,nm is compatible with 357(3)\,nm, the experimental value from $\mu$SR 
data. The Ginzburg-Landau parameter $\kappa \sim 59 \gg 1$ 
confirms once more that Re$_{0.82}$Nb$_{0.18}$ is a strong type-II superconductor.

\subsection{Transverse-Field $\mu$SR}

\begin{figure}[!htb!]
\centering
\includegraphics[width=1.0\columnwidth]{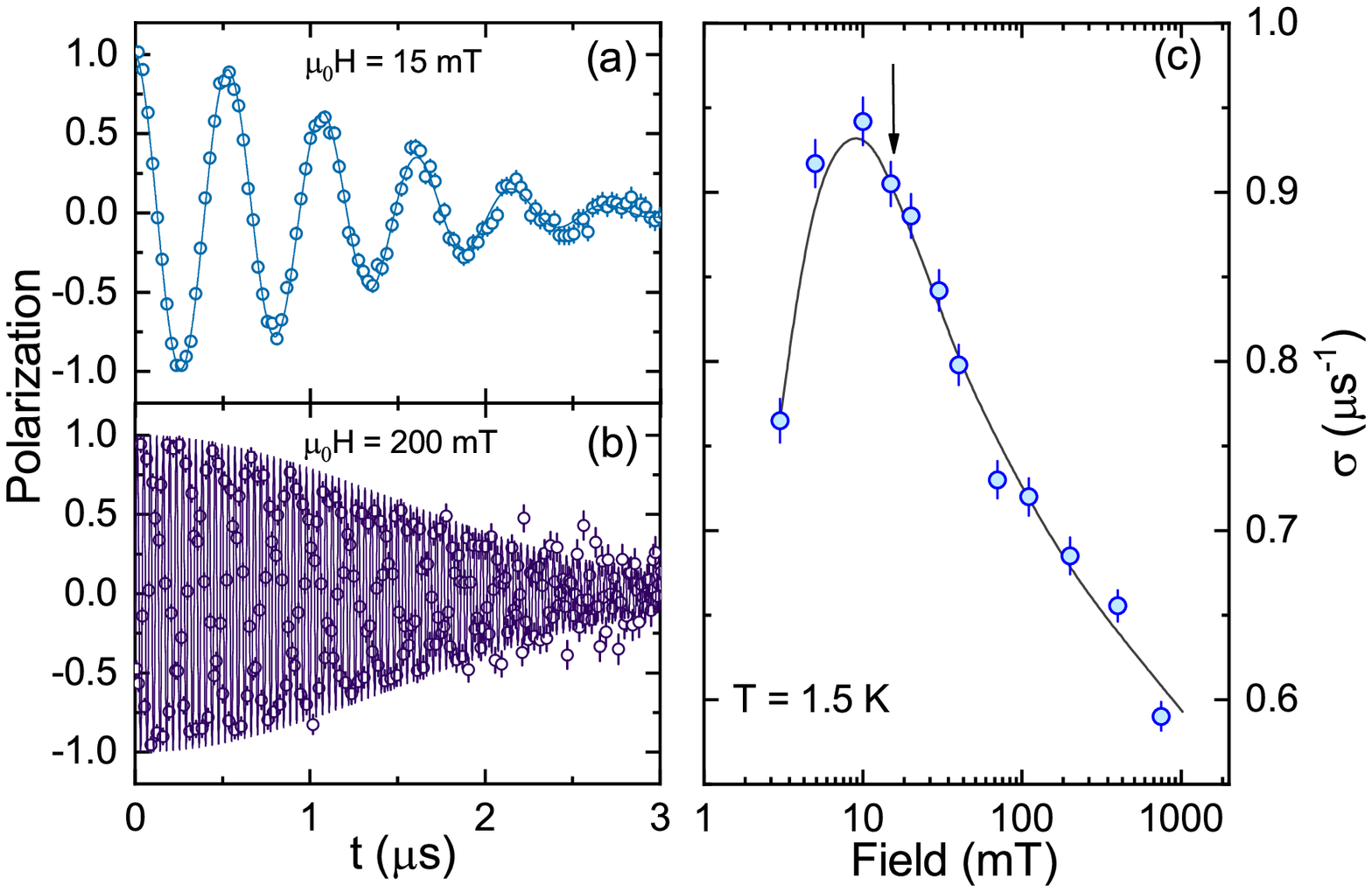} 
\caption{\label{fig:TF-muSR}Time-domain TF-$\mu$SR spectra in the superconducting 
state of Re$_{0.82}$Nb$_{0.18}$ measured at 15\,mT (a) and 200\,mT (b). 
The Gaussian relaxation rate vs.\ field exhibits a broad maximum near 10\,mT (c). 
The arrow indicates the field value used in the $T$-dependent TF-$\mu$SR 
measurements at PSI. The solid line is a guide to the eyes.}
\end{figure}

The optimal field value for TF-$\mu$SR experiments was determined via preliminary field-dependent $\mu$SR depolarization-rate measurements at 1.5 K (see Fig.~S\ref{fig:TF-muSR}). The corresponding TF spectra are described by the same model [Eq.~(1)] used in the main text.  As shown in Fig.~S\ref{fig:TF-muSR}(c), the resulting Gaussian relaxation rate $\sigma$ exhibits a maximum at 10 mT, close to the lower critical field value $\mu_{0}H_{c1}$. By considering the decrease of intervortex distance with field and the vortex-core effects, a field of 15\,mT was chosen for the temperature-dependent study at PSI, while the ISIS measurements were performed in a field of 30\,mT.
\begin{figure}[!htb]
	\centering
	\includegraphics[width=0.9\columnwidth]{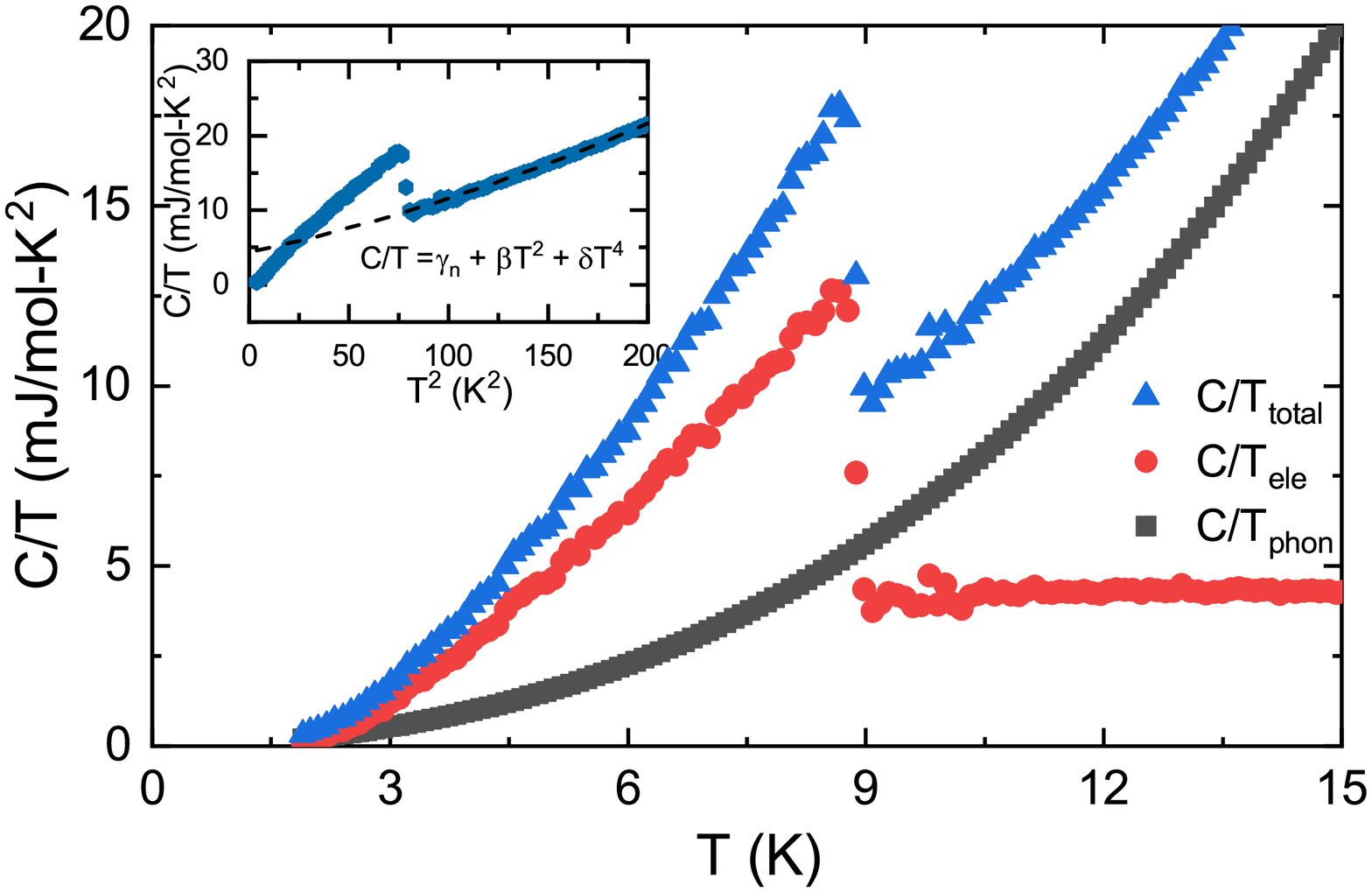} 
	\caption{\label{fig:Cp}Temperature dependence of the total, 
	electronic, and phononic specific heat of Re$_{0.82}$Nb$_{0.18}$ measured in zero field. 
	The dashed-line in the inset is a fit to $C/T = \gamma_\mathrm{n} +\beta T^2 + \delta T^4$.}
\end{figure}

\begin{table}[tbp]
	\centering
	\caption{Atomic coordinates of Re, Re$_{0.82}$Nb$_{0.18}$, Nb$_{0.5}$Os$_{0.5}$, and Mg$_{10}$Ir$_{19}$B$_{16}$. Data for 
	the last two cases were taken from Refs.~\onlinecite{Klimczuk2006,Waterstrat1977,SinghNbOs}.\label{tab:table1}} 
	\begin{ruledtabular}
		
		Re ($P6_3/mmc$; $a = b = 2.762$\,\AA, $c = 4.457$\,\AA,)\\
		\begin{tabular}{lcccc}
			\textrm{Atom}&
			\textrm{Wyckoff position}&
			\textrm{$x$}&
			\textrm{$y$}&
			\textrm{$z$}\\
			\midrule 
		Re           & 2$c$  & 0.33330  & 0.66670  & 0.25000 	\\

\end{tabular}
		\\ \vspace{10pt}
		
		Re$_{0.82}$Nb$_{0.18}$ ($I\overline{4}3m$; $a = b = c = 9.636$\,\AA)\\
		\begin{tabular}{lcccc}
			\textrm{Atom}&
			\textrm{Wyckoff position}&
			\textrm{$x$}&
			\textrm{$y$}&
			\textrm{$z$}\\
			\midrule 
			Nb$_1$            & 2$a$  & 0      & 0       & 0\\
			Nb$_2$            & 8$c$  & 0.3179 & 0.3179  & 0.3179\\
			Re$_1$            & 24$g$ & 0.3592 & 0.3592  & 0.0413\\
			Re$_2$            & 24$g$ & 0.0912 & 0.0912  & 0.2842\\ 
		\end{tabular}
		\\ \vspace{10pt}
		
		Nb$_{0.5}$Os$_{0.5}$ ($I\overline{4}3m$; $a = b = c = 9.765$\,\AA)\\
		\begin{tabular}{lcccc}
			\textrm{Atom}&
			\textrm{Wyckoff position}&
			\textrm{$x$}&
			\textrm{$y$}&
			\textrm{$z$}\\
			\midrule 
			Nb$_1$            & 2$a$   & 0      & 0       & 0\\
			Os$_1$            & 2$c$   & 0      & 0       & 0\\
			Nb$_2$           & 8$c$   & 0.3170 & 0.3170  & 0.3170\\
			Os$_2$           & 8$c$   & 0.3170 & 0.3170  & 0.3170\\
			Nb$_3$           & 24$g$ & 0.3560 & 0.3560  & 0.0420\\
			Os$_3$           & 24$g$  & 0.3560 & 0.3560  & 0.0420\\
			Nb$_4$            & 24$g$ & 0.0809 & 0.0809  & 0.2780\\
			Os$_4$            & 24$g$  & 0.0809 & 0.0809  & 0.2780\\
		\end{tabular}
		\\ \vspace{10pt}
		
		Mg$_{10}$Ir$_{19}$B$_{16}$  ($I\overline{4}3m$; $a = b = c = 10.568$\,\AA)\\
		\begin{tabular}{lcccc}
			\textrm{Atom}&
			\textrm{Wyckoff position}&
			\textrm{$x$}&
			\textrm{$y$}&
			\textrm{$z$}\\
			\midrule 
			Ir$_1$           & 2$a$    & 0      & 0       & 0\\
			Ir$_2$           & 12$d$   & 0      & 0.25    & 0.5\\
			Ir$_3$           & 24$g$   & 0.0703 & 0.2525  & 0.2525\\
			Mg$_1$           & 8$c$    & 0.3331 & 0.3331  & 0.3331\\
			Mg$_2$           & 12$e$   & 0      & 0       & 0.3473\\
			B$_1$            & 8$c$    & 0.1127 & 0.1127  & 0.1127\\
			B$_2$            & 24$g$   & 0.1639 & 0.1639  & 0.4140\\
		\end{tabular}
	\end{ruledtabular}
\end{table}

\begin{table*}[tbp]
	\centering
	\caption{Nuclear moment (in nuclear magnetons $\mu_\mathrm{N}$), muon-spin relaxation (at base temperature), calculated internal field, and superconducting parameters for $\alpha$-Mn-type Re$T$-NCSC and the pure Re metal. The internal field is calculated using $B_\mathrm{int}$ = $\sqrt{2}\sigma_\mathrm{int}/\gamma_\mu$, where $\sigma_\mathrm{int}$ represents the change of muon relaxation rate between the normal and superconducting states. Except for Re$_3$W, where $\sigma_\mathrm{int}$ is expected to be below the resolution of the $\mu$SR technique, all the listed materials show spontaneous magnetic fields below $T_c$.\label{tab:table2}} 
	\begin{ruledtabular}
		\begin{tabular}{lccccccccccccc}
			\textrm{Re$T$}&
			\textrm{$M$($\mu_\mathrm{N}$)}&
			\textrm{$\sigma$\,($\mu$s$^{-1}$)}&
			\textrm{$\Lambda$\,($\mu$s$^{-1}$)}&
			\textrm{$B_\mathrm{int}$($\mu$T)}&
			\textrm{$T_c$(K)}&
			\textrm{$\mu_0$H$_{c1}$\,(mT)}&
			\textrm{$\mu_0$H$_{c2}$\,(T)}&
			\textrm{$\xi_\mathrm{GL}$\,(\AA{})}&
			\textrm{$\lambda_\mathrm{GL}$\,(\AA{})}&
			\textrm{$\kappa_\mathrm{GL}$}&
			\textrm{$\lambda_\mathrm{eff}$\,(\AA{})}&
			\textrm{$\Delta_0$ (meV)}&
			\textrm{$\Delta C/\gamma T_{c}$}\\
			\midrule 
			Re$_{3}$W~\cite{Biswas2012,Biswas2011}            & 2.778 & 0.266 & --     & 0.0 & 7.80  & 9.7   & 12.5 & 51.3  & 2571 & 50.0   & 4180 & 1.38 & 1.50\\
			Re$_{24}$Ti$_5$~\cite{Shang2018}                  & 2.943 & 0.256 & 0.017 & 9.0 & 6.00  & 8.3   & 11.2 & 54.1  & 2860 & 53.0 & 3921 & 1.08 & 1.40\\
			Re$_{6}$Ti~\cite{Singh2018}                       & 2.990 & 0.251 & 0.027 & 14.0 & 6.00  & 5.8     & 11.5    & 53.5     & 3490   & 65.2    & 4937 & 0.95 & 1.58\\
			Re$_{6}$Zr~\cite{Singh2014,mayoh2017ReZr}         & 3.010 & 0.264 & 0.022 & 11.0 & 6.75  & 8.0   & 11.6 & 53.3  & 3696 & 69.3 & 3086 & 1.21 & 1.60\\
			Re$_{6}$Hf~\cite{Singh2017, Singh2016ReHf,chen2016ReHf}      & 2.970 & 0.262 & 0.035 & 8.5 & 5.96  & 5.6   & 12.2 & 52.0  & 3538 & 69.0 & 4620 & 1.10 & 1.53\\ 
			\midrule
			Re$_{0.82}$Nb$_{0.18}$                           & 3.910 & 0.392 & 0.007 & 38.2 & 8.8  & 6.4  & 15.6 & 45.9   & 3528 & 59.0 & 3567 & 1.61 & 1.94\\
			Re                                               & 3.207 & 0.326 & 0.003 & 17.5 & 2.7  & 3.7   & --    & --     & --    & --    & --    & --    &--    \\
		\end{tabular}
	\end{ruledtabular}
\end{table*}

\subsection{Zero-field specific heat}

Figure~S\ref{fig:Cp} shows the temperature dependence of the total, 
electronic, and phononic specific heat of Re$_{0.82}$Nb$_{0.18}$ measured 
below 15\,K in zero field. To subtract the phonon contribution from the total specific heat, the normal-state specific heat is fitted 
to the expression $C/T = \gamma_\mathrm{n} +\beta T^2 + \delta T^4$, where $\gamma_\mathrm{n}$ is the normal-state electronic specific-heat coefficient, whereas $\beta$ and $\delta$ are the phonon specific-heat coefficients~\cite{Barron1999,Tari2003}. From the fit shown in the inset of Fig.~S\ref{fig:Cp} (dashed-line), the derived $\gamma_\mathrm{n}$, $\beta$ and $\delta$ values are 4.4 mJ/mol-K$^2$, 0.0582 mJ/mol-K$^4$ and 0.00014 mJ/mol-K$^6$. The phonon contribution can hence be calculated using the expression, $C/T_\mathrm{phon} = \beta T^2 + \delta T^4$, which is then subtracted from the experimental data. The resulting $C/T_\mathrm{ele} = C/T_\mathrm{total} - C/T_\mathrm{phon}$ curve (shown with red symbols in Fig.~S\ref{fig:Cp}), represents the electronic specific heat. 
\begin{figure}[!htb]
	\centering
	\includegraphics[width=1\columnwidth]{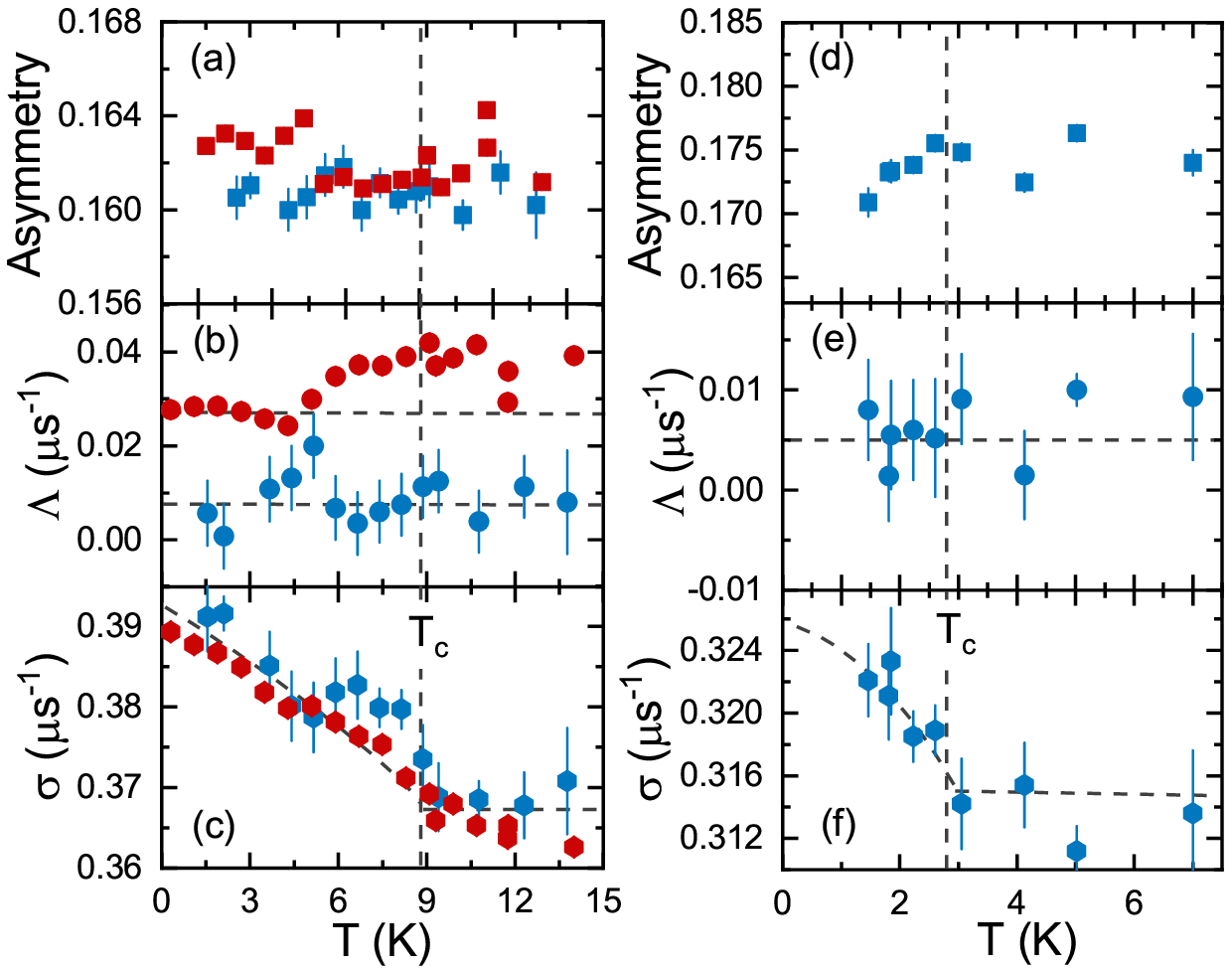} 
	\caption{\label{fig:ZFlamfree}ZF-$\mu$SR fit parameters vs.\ temperature for Re$_{0.82}$\-Nb$_{0.18}$ (a)-(c) 
	and pure Re (d)-(f). Blue symbols represent datasets collected at PSI, while red symbols refer to datasets collected 
	at ISIS. For comparison, the asymmetry of data collected at ISIS was divided by a factor of 1.25. The dashed lines are guide to the eyes.}
\end{figure}
\subsection{Zero-field $\mu$SR}
\begin{figure}[!htb]
	\centering
	\includegraphics[width=1\columnwidth]{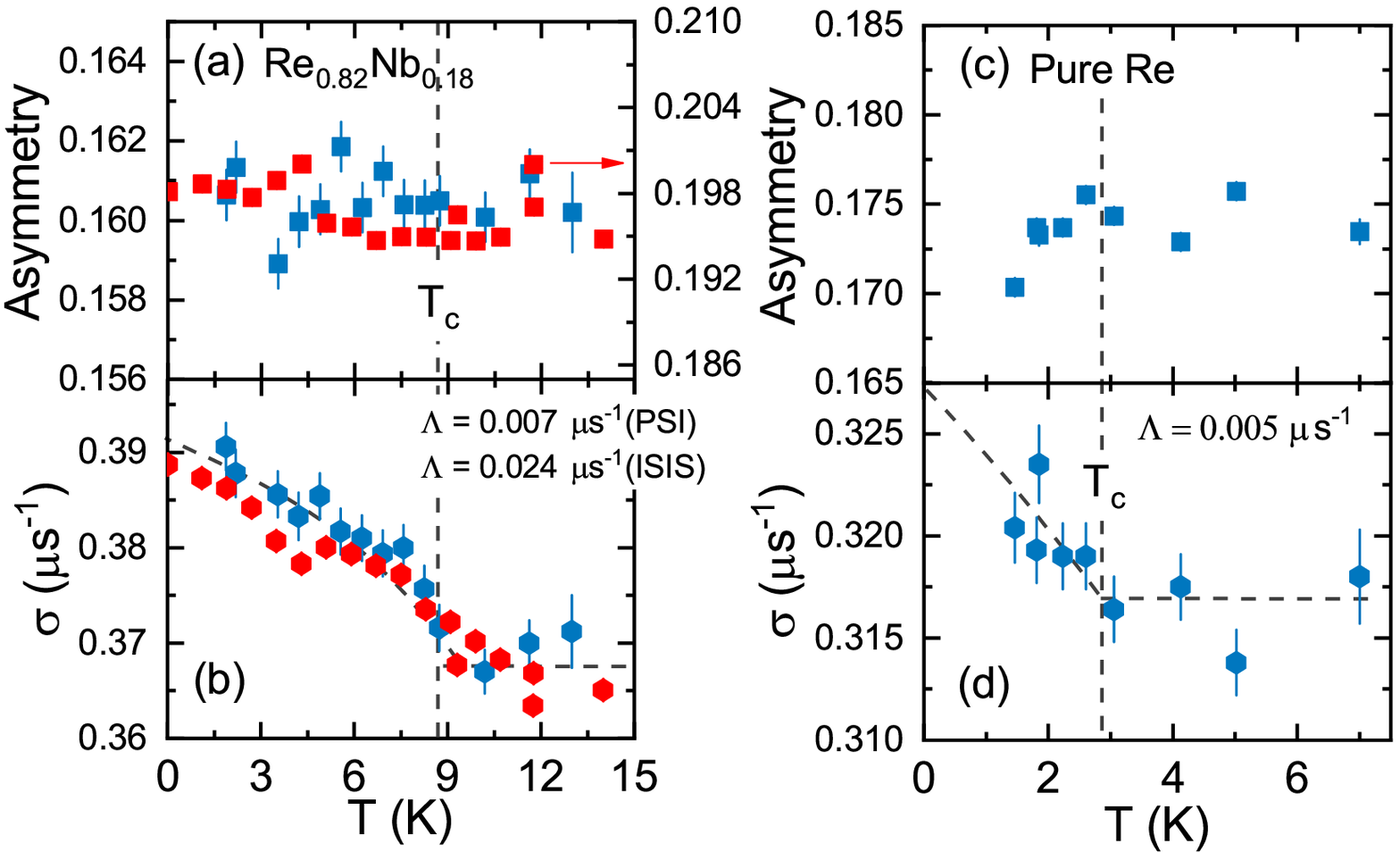} 
	\caption{\label{fig:ZFlamfix}ZF-$\mu$SR fit parameters vs.\ temperature for Re$_{0.82}$Nb$_{0.18}$ (a)-(b) and pure Re (c)-(d). The reported $\Lambda$ value for Re$_{0.82}$Nb$_{0.18}$ is the average of 0.007 (PSI) and 0.024 (ISIS), while for Re it is 0.005 $\mu$s$^{-1}$. The blue symbols refer to datasets collected at PSI, while the red symbols to datasets collected at ISIS. The dashed lines are guide to the eyes.}
\end{figure}
The ZF-$\mu$SR data were analyzed by combined Lorentzian and Gaussian Kubo-Toyabe relaxation function [see details in Eq.~(3) in the main text]. The polarization parameters $P_\mathrm{s}$ and $P_\mathrm{bg}$ were obtained by dividing the time-dependent asymmetry $A(t)$ by its initial value $A_0$. The initial asymmetry parameter $A_0$ was allowed to change with temperature, while the background signal was fixed for all the temperatures since, considering the limited temperature range of the measurements, we do not expect an appreciable change in background. The initial asymmetry of the background signal $A_\mathrm{bg}$ ($P_\mathrm{bg}$ = $A_\mathrm{bg}$/$A_0$) was determined by fits to data at base temperature and then fixed for all the temperatures. It is close to 3\% of total asymmetry $A_0$ in Re$_{0.82}$Nb$_{0.18}$ (PSI) and to 2\% in pure Re. The reduced total asymmetry for datasets collected at PSI [see Fig.~S\ref{fig:ZFlamfree}(a)] reflects ZF-measurements carried out in transverse mode, i.e., with the initial muon-spin perpendicular to muon momentum, implying a loss of one third of the asymmetry. From Fig.~S\ref{fig:ZFlamfree}, only the Gaussian muon-relaxation rate $\sigma(T)$ exhibits a small but distinct increase below $T_c$, while the other parameters show a weak temperature dependence with no clear anomaly across $T_c$. In case of pure Re powders, the data in the inset of Fig.~3(b) of the main text were fitted with a fixed $\Lambda$ value of 0.003~$\mu$s$^{-1}$. Here we re-analyze the ZF-$\mu$SR data on Re powders by releasing $\Lambda$. The derived parameters are summarized in Figs.~S\ref{fig:ZFlamfree} (d)-(f); again only $\sigma(T)$ exhibits clear anomaly below $T_c$. 

\begin{figure}[!htb]
	\centering
	\includegraphics[width=0.8\columnwidth]{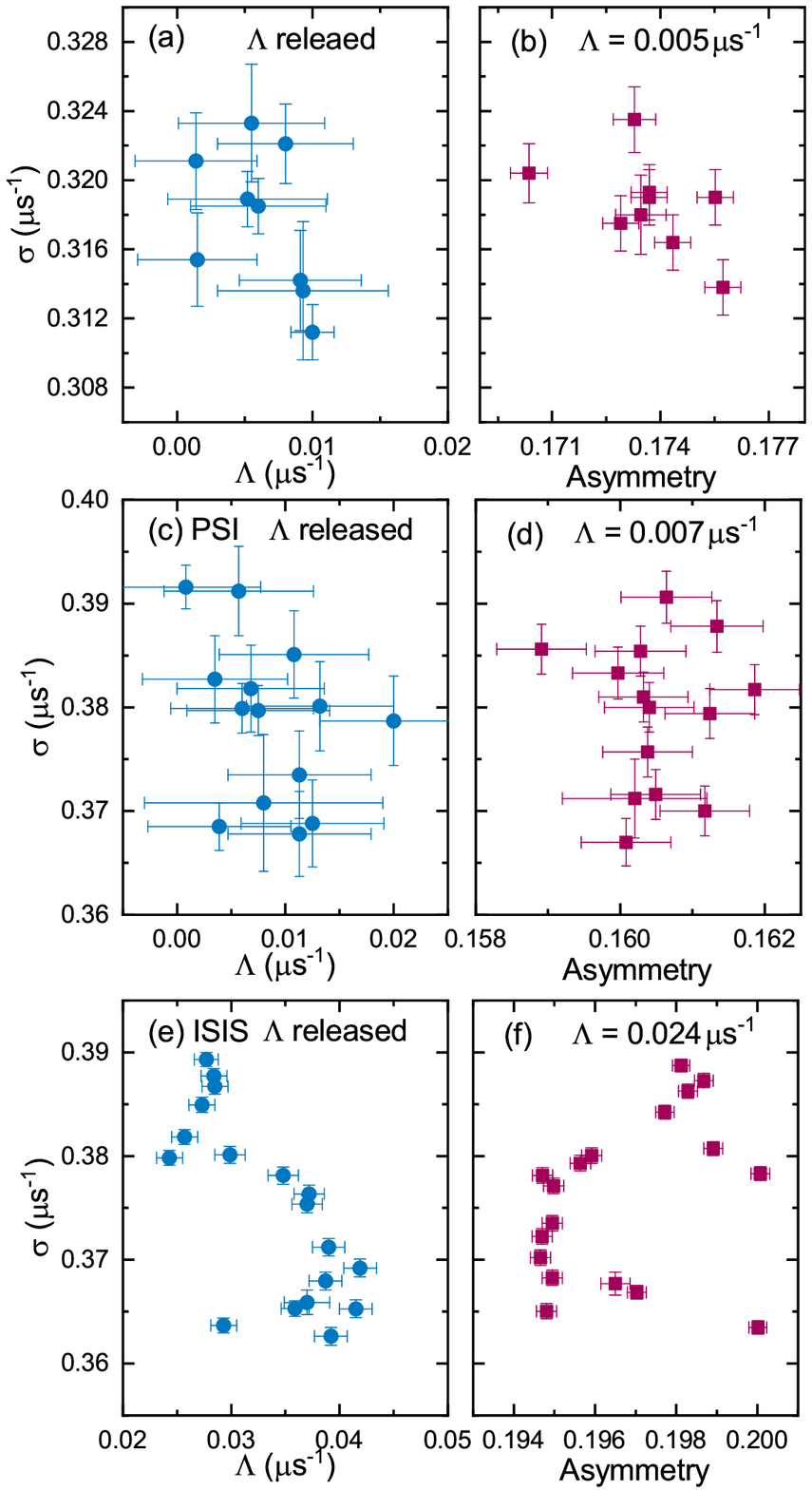} 
	\caption{\label{fig:correlation}Cross correlations of ZF-$\mu$SR fit parameters for pure Re (a)-(b) and Re$_{0.82}$Nb$_{0.18}$ (c)-(f). The vertical axis shows $\sigma$, while the horizontal axis $\Lambda$ or asymmetry. In both cases, the rather random distribution of points indicates a lack of any relevant correlations between the fit parameters.}
\end{figure}

To confirm the above conclusions, all ZF-$\mu$SR data were re-analyzed by fixing $\Lambda$s to the respective average values, as estimated from preliminary analyses with freely varying $\Lambda$ values (see Fig.~S\ref{fig:ZFlamfree}). As shown by dashed lines in pannels (b) and (e), the average $\Lambda$ values turned out to be 0.007 (PSI)/0.024 (ISIS) and 0.005~$\mu$s$^{-1}$ for Re$_{0.82}$Nb$_{0.18}$ and pure Re, respectively. In both cases, the derived parameters, including the asymmetry and muon-spin relaxation rate $\sigma$, are shown in Fig.~S\ref{fig:ZFlamfix}. Clearly, the evolution of $\sigma$ with temperature remains robust in both compounds, hence implying that the spontaneous magnetic fields we detect via ZF-$\mu$R are an \emph{intrinsic effect} and not a fit artifact.
This is further confirmed in Fig.~S\ref{fig:correlation}, where we show the cross correlations between the different fit parameters. The lack of any visible trends (i.e., the lack of correlations) confirms once more the \emph{intrinsic nature of TRSB} in both pure Re and Re$_{0.82}$Nb$_{0.18}$.

\subsection{Symmetry analysis}
\tcb{
In this section we \tcr{provide further} details \tcr{on} the symmetry-allowed 
TRS-breaking order \tcr{parameter in} 
Re$_{0.82}$Nb$_{0.18}$ and \tcr{in pure} Re.}
\subsubsection{Re$_{0.82}$Nb$_{0.18}$} 
\begin{figure*}[!htb]
\centering
\includegraphics[width=0.97\textwidth]{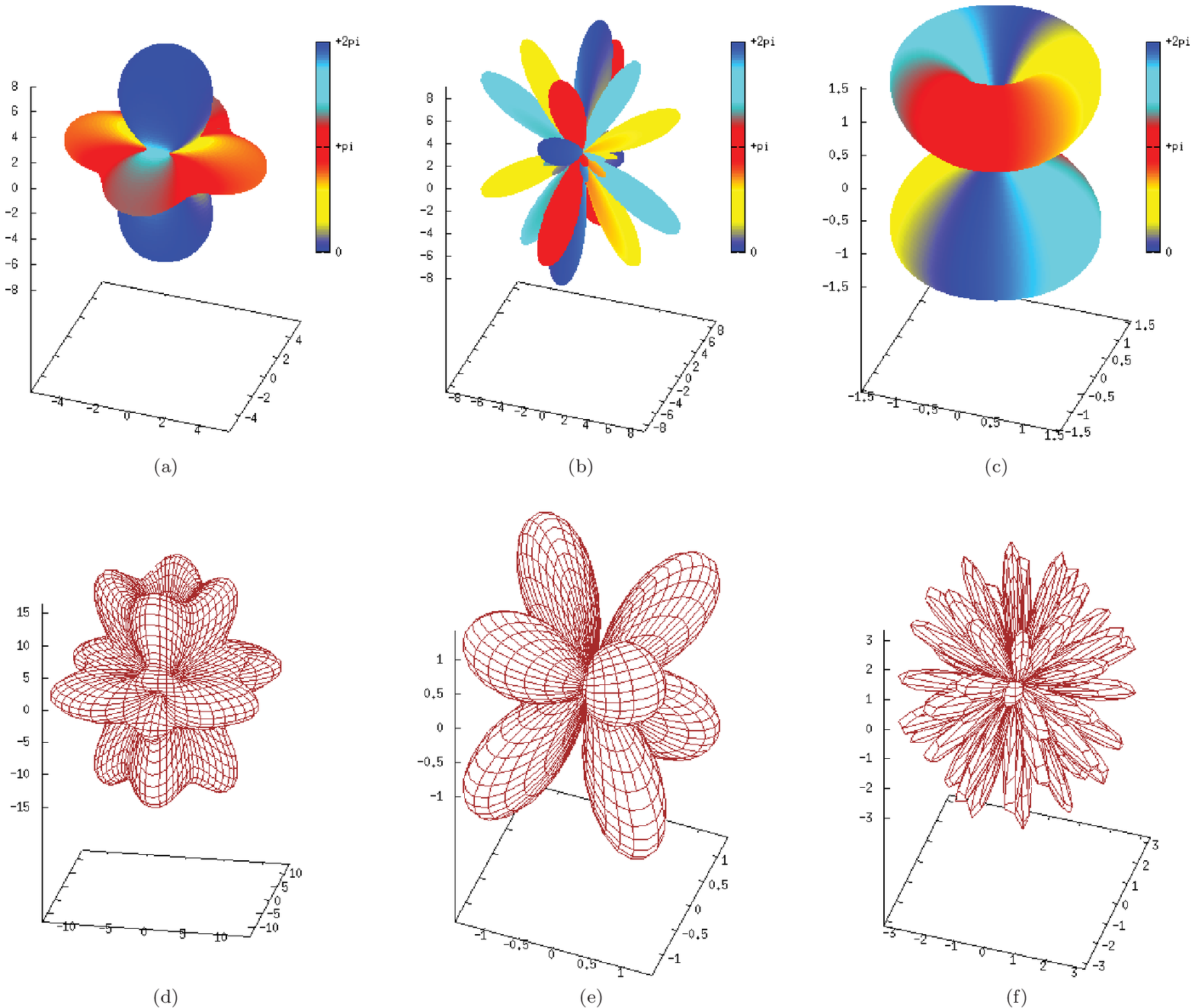} 
\caption{Polar plot of the \tcr{symmetry-allowed} TRSB order parameters 
\tcr{in} Re$_6$T, \tcr{for the $E$, $F_1$, and $F_2$ irreps, respectively.} 
Panels (a), (b), and (c) depict the \tcr{\emph{magnitude}} of the \tcr{\emph{singlet}} 
part of the order parameter, with the corresponding \tcr{\emph{phase}} shown 
in color. Panels (d), (e), and (f) \tcr{instead} show the magnitude of the 
 \tcr{\emph{triplet}} order parameter $\vec{d}(\vec{k})$ for the three cases, 
 respectively.}
\label{fig:Re6T_order_para}
\end{figure*}

\tcb{The Re$_{0.82}$Nb$_{0.18}$ system has the same non-cen\-tro\-sym\-met\-ric{} 
\tcr{$bcc$} crystal structure as \tcr{other} Re$_6$(Zr, Hf, Ti) materials 
\tcr{with a} symmorphic space group \tcr{$I\bar{4}3m$} \tcr{and} corresponding 
point group \tcr{$T_d$}. \tcr{A} standard symmetry analysis~\cite{Singh2014} 
finds that both the 2D irreducible representation (irrep) and the two 3D 
irreps of $T_d$ support TRS breaking instabilities. Assuming strong SOC and 
writing the pair potential as 
$\hat{\Delta}(\vec{k}) = i \left[ \Delta_0(\vec{k}) + \vec{d}(\vec{k}).\hat{\sigma}\right]\hat{\sigma}_y$, 
the TRS breaking instability for the 2D irrep corresponds to~\cite{Singh2014}: 
\begin{eqnarray}\nonumber
\Delta_0(\vec{k}) & = & (2 k_z^2 - k_x^2 - k_y^2) + i(k_x^2 - k_y^2), \\
\vec{d}(\vec{k}) & = & [k_x(k_y^2 - k_z^2), k_y(k_x^2 - k_z^2), k_z(k_x^2 - k_y^2)] \Delta_0(\vec{k}).\nonumber
\end{eqnarray}
This leads to a vanishing pair potential at eight symmetry-required point \tcr{nodes,}  
given by $\phi = \pm \pi/4, \pm 3\pi/4$ and $\theta = \arctan \left(\sqrt{2}\right),\pi-\arctan \left(\sqrt{2}\right)$, 
as shown in Fig.~S\ref{fig:Re6T_order_para}(a) and S\ref{fig:Re6T_order_para}(d).} 

\tcb{Similarly, for the \tcr{two 3D irreps,} $F_1$ and $F_2$, the TRS 
breaking pair potentials are, { respectively}~\cite{Singh2014}:
\begin{eqnarray}\nonumber
\Delta_0(\vec{k}) & = & k_z(k_y + i k_x) (k_x^2 - k_y^2)(k_y^2 - k_z^2) (k_z^2 - k_x^2), \\
\vec{d}(\vec{k}) & = & (1,i,0) k_x k_y k_z \nonumber
\end{eqnarray}
and
\begin{eqnarray}
\Delta_0(\vec{k}) & = & (k_y + i k_x)k_z, \nonumber\\
\vec{d}(\vec{k}) & = & (1,i,0) k_x k_y k_z(k_x^2 - k_y^2)(k_y^2 - k_z^2) (k_z^2 - k_x^2).\nonumber
\end{eqnarray}
In both \tcr{cases,} the pair potential \tcr{vanishes} at the `North' 
and `South' poles, $\theta = 0$ and $\pi$, and \tcr{at the} line nodes 
at the `equator', $\theta=\pi/2$, as shown in Fig.~S\ref{fig:Re6T_order_para}(b) 
and S\ref{fig:Re6T_order_para}(e) for the $F_1$ case and in Fig.~S\ref{fig:Re6T_order_para}(c) 
and S\ref{fig:Re6T_order_para}(f) for the $F_2$ case.}

\begin{figure*}[!htb]
\centering
\includegraphics[width=0.97\textwidth]{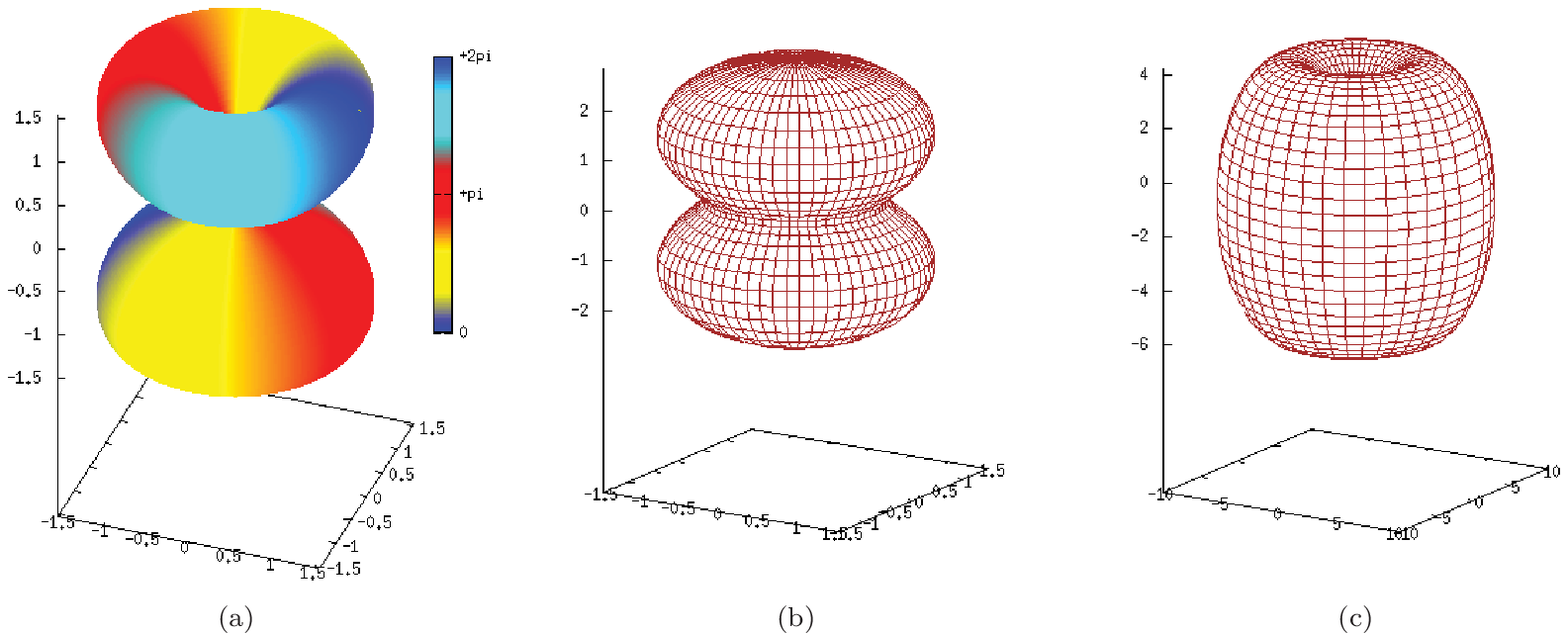} 
\caption{Polar plot of the TRSB order parameters of pure Re corresponding 
to the $D_{3d}$ subgroup (taking $A=1/2$ and $B=1/3$). (a) The 
\tcr{\emph{magnitude}} of the \tcr{\emph{singlet}} TRSB order parameter 
with the corresponding \tcr{\emph{phase}} shown in color. (b) The magnitude 
of $\vec{d}(\vec{k})$ corresponding to the \tcr{\emph{triplet}} TRSB order 
parameter. (c) The lowest excitation gap at the Fermi energy for the 
triplet TRSB order parameter given in Eq.~(S\ref{eqn:gap}).}
\label{fig:Re_order_para}
\end{figure*}

\subsubsection{\tcr{Pure} Re}
\tcb{Re has a hexagonal structure with nonsymmorphic space group \tcr{$P6_3/mmc$} 
(\tcr{No.\ 194 --} point group \tcr{$D_{6h}$}) 
with two atoms per unit cell. Due to its centrosymmetric nature, this 
space group has separate singlet and triplet uniform superconducting 
instabilities. \tcr{They} are given by the symmetries of the factor group 
$\mathcal{G}_{194} = $\{$P6_3/mmc$\}$/\mathcal{T}$ ($\mathcal{T}$ is the 
group of pure primitive translations), which is isomorphic to \tcr{$D_{6h}$}. 
Hence, $\mathcal{G}_{194}$ has the 2D irreps $E_g$ and $E_u$ which, in 
principle, allow the TRS breaking \tcr{of the order parameter} by a nontrivial 
phase difference between the two components of a possible two-component 
order parameter. In general, \tcr{compared to a symmorphic crystal,} 
finding the possible forms of the gap function is complicated by the fact 
that the basis functions of the point group $D_{6h}$ cannot \tcr{simply be} 
used as building blocks of the pairing potential. \tcr{Nevertheless, it is} 
straightforward to enumerate the pairing potentials that do not break 
glide-plane or screw-axis symmetries. These correspond to the 
\tcr{$D_{3d}$ subgroup} of $\mathcal{G}_{194}$, containing only its 
symmorphic symmetries. We note that $D_{3d} = D_3 \otimes i$, which is 
the group of symmetries of an equilateral triangle. The simplest basis 
functions of $D_3$ are given in Table S\ref{tab:D3}. 
\tcr{Clearly,}  
$D_3$ does include \tcr{a 2D} irrep and thus allows for TRS breaking 
both in the \tcr{singlet-} ($E_g$ irrep of $D_{3d}$) or \tcr{in} the 
triplet channel ($E_u$ irrep of $D_{3d}$).} 
%

{\renewcommand{\arraystretch}{1.10}{%
\begin{table}
 \centering
  \begin{ruledtabular}
\caption{\label{tab:D3}Basis functions of $D_3$.} 
\begin{tabular}{ccc} 
$D_3$ & \multicolumn{2}{c}{Basis functions}\\ 
Irreps & Scalar (even) & Vector (odd) \\
\midrule
$A_1$ & $\left[A + B (k_x^2 + k_y^2) + C k_z^2 \right]$ & $\left[ A(k_x \hat{x} + k_y \hat{y}) + B k_z \hat{z} \right]$ \\ 
$A_2$ & $A k_z k_x (k_x^2 - 3 k_y^2)$ & $A k_y (k_y^2 - 3 k_x^2) \hat{z}$ \\ 
$E$   & $Ak_z \left(\begin{array}{c} k_x \\ k_y\end{array}\right)$ & $\left(\begin{array}{c} A k_z \hat{x} + B k_x \hat{z} \\ A k_z \hat{y} + B k_y \hat{z}\end{array}\right)$ \\ 
	\end{tabular}
	\end{ruledtabular}
\end{table}
}

In this case, to compute the possible singlet ground states, 
we write the most general form of the order parameter just below $T_c$,  
compatible with the relevant basis functions:  
\begin{equation}
\Delta_0(\boldsymbol{k}) = (\eta_1 k_z k_x + \eta_2 k_z k_y)
\end{equation}
and then obtain a general expression of the free energy, expressed as a 
function of the two pairing components $\eta_1,\eta_2$, by requiring 
that it respects the symmetries of $D_{3d}$: 
\begin{equation}\nonumber
\mathcal{F} = a (|\eta_1|^2 + |\eta_2|^2) + \beta_1 (|\eta_1|^2 + |\eta_2|^2)^2 + \beta_2 (\eta^*_1 \eta_2 - \eta_1 \eta^*_2)^2 + \ldots
\end{equation}
Minimization of this quartic polynomial gives two stable ground states: $(\eta_1,\eta_2) = (1, 0)$ and $(1, i)$. The latter is the only one that breaks TRS and it gives
\begin{equation}
\Delta_0(\vec{k}) = A k_z(k_x + i k_y).
\end{equation} 
This corresponds to a singlet energy gap $|A| |k_z| \sqrt{k^2_x + k^2_y}$. 
The \tcr{low-energy} excitations would thus be dominated by a line node 
at the `equator', $\theta=\pi/2$, and point nodes at $\theta=0, \pi$, 
as shown in Fig.~S\ref{fig:Re_order_para}(a).

\tcb{A similar procedure applied to the triplet channel gives one TRS 
breaking state with \tcr{a $d$-}vector: 
\begin{equation}
\vec{d}(\vec{k}) = \left[A k_z, i A k_z, B(k_x + i k_y) \right].
\end{equation}
The variation of its magnitude is shown in Fig. S\ref{fig:Re_order_para}(b). 
The corresponding lowest excitation energy gap is given by: 
\begin{equation}\label{eqn:gap}
\tcr{\Delta_\mathrm{min}} = \sqrt{g(k_x,k_y) + 2 A^2 k^2_z  - 2 | A |k_z| \sqrt{g(k_x,k_y) +  A^2 k^2_z }}
\end{equation}
where $g(k_x,k_y) = B^2 (k^2_x + k^2_y)$. 
\tcr{As shown in Fig.~S\ref{fig:Re_order_para}(c),} it has two point nodes 
at the `poles', $\theta=0,\pi$, \tcr{but} no line nodes.}


%

\end{document}